\documentclass[floatfix,twocolumn,showpacs,aps,tightenlines,superscriptaddress]{revtex4-1}
\usepackage{graphicx}
\usepackage{color}
\usepackage{psfrag}
\usepackage{dcolumn}
\usepackage{bm}

\usepackage{amssymb}
\usepackage{amsmath}
\usepackage{amsfonts}
\usepackage{longtable}
\usepackage{xspace}
\usepackage{verbatim}

\newcommand{\beq}{\begin{equation}}
\newcommand{\eeq}{\end{equation}}

\newcommand{\be}{\begin{equation}}
\newcommand{\ee}{\end{equation}}
\newcommand{\bea}{\begin{eqnarray}}
\newcommand{\eea}{\end{eqnarray}}
\newcommand{\bes}{\begin{subequations}}
\newcommand{\ees}{\end{subequations}}

\newcommand{\MPA}{{moving punctures approach}\xspace}
\newcommand{\hispid}{{\sc HiSpID}\xspace}

\usepackage{bm}

\begin{document}

\title{Evolutions of unequal mass, highly spinning black hole binaries}

\author{James Healy} 
\author{Carlos O. Lousto}
\affiliation{Center for Computational Relativity and Gravitation,
School of Mathematical Sciences,
Rochester Institute of Technology, 85 Lomb Memorial Drive, Rochester,
 New York 14623}
\author{Ian Ruchlin} 
\affiliation{Department of Mathematics, West Virginia University,
Morgantown, West Virginia 26506, USA}
\affiliation{Center for Computational Relativity and Gravitation,
School of Mathematical Sciences,
Rochester Institute of Technology, 85 Lomb Memorial Drive, Rochester,
 New York 14623}
\author{Yosef Zlochower} 
\affiliation{Center for Computational Relativity and Gravitation,
School of Mathematical Sciences,
Rochester Institute of Technology, 85 Lomb Memorial Drive, Rochester,
 New York 14623}

\date{\today}

\begin{abstract}
We evolve a binary black hole system bearing a mass ratio of $q=m_1/m_2=2/3$
and individual spins of $S^z_1/m_1^2=0.95$ and $S^z_2/m_2^2=-0.95$ in a
configuration where the large black hole has its spin antialigned with the orbital
angular momentum, $L^z$, and the small black hole has its spin aligned with $L^z$.
This configuration was chosen to measure the maximum recoil of the remnant
black hole for nonprecessing binaries. We find that the remnant black hole
recoils at  $500km/s$, the largest recorded value from numerical simulations 
for aligned spin configurations. The remnant mass, spin, and
gravitational waveform peak luminosity and frequency also provide a valuable
point in parameter space for source modeling.
\end{abstract}

 \maketitle

\section{Introduction}\label{sec:intro}

Since the breakthroughs in numerical relativity of
2005~\cite{Pretorius:2005gq, Campanelli:2005dd, Baker:2005vv} it is
possible to accurately simulate
moderate-mass-ratio and moderate-spin black-hole
binaries. State of the art numerical relativity
codes now routinely evolve
binaries with mass ratios as small as $q\lesssim
1/16$~\cite{Gonzalez:2008bi, Lousto:2010qx, Lousto:2010ut,
Sperhake:2011ik, Chu:2015kft, Jani:2016wkt}, and are pushing towards
much smaller mass ratios. Indeed, there have been
some initial explorations
of $q=1/100$ binaries~\cite{Lousto:2010ut, Sperhake:2011ik}.

However, when it comes to highly-spinning binaries,
prior to the work of~\cite{Lovelace:2008tw} of the SXS
Collaboration \footnote{{\tt https://www.black-holes.org}}, it
was not even possible to construct initial data for binaries with
spins larger than~$\sim 0.93$~\cite{Cook:1989fb}. This limitation was
due to the use of conformally flat initial data. Conformal flatness is
a convenient assumption because the Einstein constraint system takes on a
particularly simple form. Indeed, using the puncture approach, the
momentum constraints can be solved exactly using the Bowen-York
ansatz~\cite{Bowen:1980yu}. There were several attempts to increase
the spins of the black holes while still preserving conformal
flatness~\cite{Dain:2002ee, Lousto:2012es}, but these introduced
negligible improvements. Lovelace {\it et al.}~\cite{Lovelace:2008tw}
were able to overcome these limitations by choosing the initial data
to be a superposition of conformally Kerr black holes in the
Kerr-Schild gauge. Using these
new data, they were able to evolve binaries with spins as large
as
0.97~\cite{Lovelace:2011nu} and, later, spins as high as
0.994~\cite{Scheel:2014ina}. Production
simulations remain still very lengthy.


Recently, we introduced a version of
highly-spinning initial data, also based on the superposition of two
Kerr black holes~\cite{Ruchlin:2014zva, Healy:2015mla}, but this time
in a puncture gauge. The main differences between the two approaches
is how easily the latter can be incorporated into moving-punctures
codes. In Refs.~\cite{Ruchlin:2014zva,Zlochower:2017bbg}, 
we were able to evolve an
equal-mass binary with aligned spins, and spin magnitudes of
$\chi=0.95$ and $\chi=0.99$ respectively, using this new data and compare 
with the results of the Lovelace {\it et al.}, finding excellent
agreement.

Studies of aligned spin binaries have provided insight on the 
basic spin-orbit dynamics of black hole mergers and also allow
for a first approximation for source parameter estimations of gravitational 
wave signals~\cite{Lange:2017wki}
because this reduced parameter space~\cite{Healy:2017psd} contains two of
the most important parameters for the modeling waveforms: 
the mass ratio (in addition to the total mass) and the spin 
components along the orbital angular momentum~\cite{Campanelli:2006uy}.

In \cite{Healy:2014yta} we found, after extrapolation
of a fitting formula, that the maximum recoil for binaries with  aligned/anti-aligned spins
occurs when the mass ratio between the smaller and larger black hole
is near $q=2/3$.
Since that study used
Bowen-York initial data, we were  not been able to produce actual simulations
of near-maximal spinning holes to verify this prediction. In this
paper, we revisit this configuration with
our new \hispid initial data, which is able to generate binaries with
spins much closer to unity. Here we evolve a binary with spins 
$\chi_i=0.95$ and measure a recoil of $\sim500km/s$, the largest
recoil ever obtained for such nonprecessing binary black hole mergers.

In this paper, we show the results of a simulation of unequal-mass
binary with aligned spins of $\chi=0.95$. There is no similar simulation
to our knowledge in the literature, thus filling a gap in the 
gravitational waveforms template bank to be used in gravitational
wave observations.
Indeed, another area of interest is the use of numerical relativity waveforms
in the detection and parameter estimation of gravitational wave signals
as observed by LIGO and other detectors \cite{Abbott:2016apu,Lange:2017wki}.
This important region of
parameter space of highly spinning binaries is currently poorly covered
by current catalogs~\cite{Mroue:2013xna, Jani:2016wkt, Healy:2017psd}
and benefits from new, accurate simulations.

We use the following standard conventions throughout this paper.
In all cases, we use geometric units where $G=1$ and $c=1$. 
Latin letters ($i$, $j$, $\ldots$) represent spatial indices.
Spatial 3-metrics are denoted by $\gamma_{ij}$ and extrinsic
curvatures by $K_{ij}$. The trace-free part of the extrinsic curvature
is denoted by $A_{ij}$. A tilde indicates a conformally related
quantity. Thus $\gamma_{ij} = \psi^4 \tilde \gamma_{ij}$ and $A_{ij} =
\psi^{-2} \tilde A_{ij}$, where $\psi$ is some conformal factor. We
denote the covariant derivative associated with $\gamma_{ij}$ by $D_i$
and the covariant derivative associated with $\tilde \gamma_{ij}$ by
$\tilde D_i$. A lapse function is denoted by $\alpha$, while a shift
vector by $\beta^i$.

This paper is organized as follows. In Sec.~\ref{sec:ID}, we provide a
brief overview of how the initial data are constructed. In
Sec.~\ref{sec:evolution} we describe the numerical techniques used to
evolve these data. In Sec.~\ref{sec:results}, we present detailed
waveform, trajectories, masses and spin results of the binary evolution. In
Sec.~\ref{sec:diagnostic}, we analyze the various diagnostics to
determine the accuracy of the simulation. 
We also provide values for the final remnant mass, spin and recoil 
velocity as well as the peak luminosity and corresponding peak 
frequency as derived from the gravitational waveform.
Finally, in
Sec.~\ref{sec:discussion}, we discuss our results on the light of
applications to parameter estimation and follow up simulations to
gravitational wave observations.

\section{Numerical Techniques}\label{sec:techniques}

\subsection{Initial Data}\label{sec:ID}
We construct initial data for a black-hole binary with individual
spins $\chi_{1,2} = 0.95$ using the \hispid 
code~\cite{Ruchlin:2014zva, Healy:2015mla}, with the modifications
introduced in~\cite{Zlochower:2017bbg}. The \hispid code solves
the four Einstein constraint equations using the   conformal
transverse traceless
decomposition~\cite{York99, Cook:2000vr, Pfeiffer:2002iy,
AlcubierreBook2008}. 

In this approach, the spatial metric
$\gamma_{ij}$ and extrinsic curvature $K_{ij}$ are given by
\begin{eqnarray}
  \gamma_{ij} = \psi^4 \tilde \gamma_{ij},\\
  K_{ij} = \psi^{-2}\tilde A_{ij} + \frac{1}{3} K \gamma_{ij},\\
  \tilde A_{ij} = \tilde M_{ij} + (\tilde{\mathbb{L}} b)_{i j},
\end{eqnarray}
where the conformal metric $\tilde \gamma_{ij}$, the trace of the
extrinsic curvature $K$, and the trace-free tensor $\tilde M_{ij}$ are
free data. The Einstein constraints then become a set of four coupled
elliptical equations for the scalar field $u = \psi - \psi_0$ and components of the spatial vector
$b^i$ ($\psi_0$ is a singular function specified analytically).
The resulting elliptical equations are solved using an
extension to the \textsc{TwoPunctures}~\cite{Ansorg:2004ds} thorn.

%
To get $\tilde \gamma_{ij}^{(\pm)}$, etc.,  we
start with Kerr black holes in quasi-isotropic (QI) coordinates and perform
a fisheye (FE) radial coordinate transformation followed by a Lorentz
boost (see~\cite{Zlochower:2017bbg} for more details).
The FE transformation is needed because it expands the horizon size,
which greatly speeds up the convergence of the elliptic solver and has
the form
\begin{equation}
    r_{\rm QI} = r_{\rm FE} [1-A_{\rm FE} \exp(-r_{\rm FE}^2/{s_{\rm
    FE}}^2)],
  \end{equation}
  where $r_{\rm FE}$ is the fisheye radial coordinate, $r_{\rm QI}$ is
  the original QI radial coordinate, and  $A_{\rm FE}$ and $s_{\rm
  FE}$ are parameters.

We use
an attenuation function described in~\cite{Ruchlin:2014zva, Zlochower:2017bbg} to modify both the
metric and elliptical equations inside the horizons,
where the attenuation function $g$ takes the form
\begin{align*}
g &= g_{+}\times g_{-} \; ,\\
  g_{\pm} &= 
          \begin{cases} 
     1 & \mbox{if } r_\pm > r_{\rm max} \\
     0 & \mbox{if } r_\pm < r_{\rm min} \\
            {\cal G}(r_{\pm}) & \mbox{otherwise},
  \end{cases} \; ,\\
  {\cal G}(r_\pm) &= \frac{1}{2}\left[1+ \tanh\left(\tan\left[ \frac{\pi}{2}
  \left(-1 + 2 \frac{r_{\pm}-r_{\rm min}}{r_{\rm max} -
  r_{\rm min}}\right)\right]\right)\right],
\end{align*}
$r_{\pm}$ is the coordinate distance to puncture $(+)$ or
$(-)$,
and the parameters $r_{\rm min} < r_{\rm max}$ are chosen
to be within the horizon. 

Finally, far from the holes, we attenuate  $\tilde \gamma_{ij}$, $K$,
and $\psi_0$. This is achieved by consistently changing the metric
fields
and their derivatives so that
\begin{eqnarray}
  \tilde \gamma_{ij}^{(\pm)} \to f(r_\pm) (\tilde \gamma_{ij}^{(\pm)} - \delta_{ij}) +
  \delta_{ij},\\ 
  K^{(\pm)} \to f(r_\pm) K^{(\pm)},\\
\left(\psi_{(\pm)}-1\right) \to f(r_\pm) \left(\psi_{(\pm)}-1\right),
\end{eqnarray}
where $f(r) = \exp(-r^4/s_{\rm far}^4)$ and $r_{\pm}$ is the
coordinate distance to puncture $(+)$ or $(-)$.
                                                                    
For compatibility with the original {\sc TwoPunctures} code, we chose
to set up \hispid so that the parameters of the binary are specified
in terms of momenta and spins of the two holes. However, unlike for
Bowen-York data, the values specified are only approximate, as the
solution vector $b^i$ can modify both of these. In practice, we find
that the spins are modified by only a trivial amount while
orbital angular momentum (as measured from the difference between the
ADM angular momentum and the two spin angular momenta) is reduced significantly. Furthermore, for
this unequal-mass case (and generally when the two black holes are not
identical), the linear momentum of the two black holes are modified by
different amounts. This means that the system with the default
parameters will have net ADM linear momentum. To compensate for both
of these changes, the boost applied to each black hole needs to
be adjusted. In practice, the change in orbital angular momentum is
the larger of the two. We adjust these boosts using an iterative
procedure. To compensate for the missing angular momentum, we increase
the magnitude of the linear momentum of each black hole by a factor of $\delta
L/D$, where $\delta L$ is the {\it missing} angular momentum and $D$
is the separation of the two black holes in quasi-isotropic coordinates. This
process is repeated until the orbital angular momentum is within 1 part
in 10 000 of the desired value. To remove excess linear momentum, we
subtract half the measured net linear momentum from each black hole. Here, we
repeat this subtraction until the measured linear momentum is smaller
than $10^{-6}M$. The net effect is that the two black holes have linear
momentum parameters with different magnitudes, and both black holes have
linear momentum parameters larger in magnitude
than those predicted by simple quasicircular conditions would
imply~\cite{Healy:2017zqj}. All parameters for the $\chi=0.95$ run are
given in Table~\ref{tab:id}.
Finally, in order to get a satisfactory solution for the initial
data problem, we used $450\times450\times22$ collocation points (the
third dimension is an axis of approximate symmetry).

\begin{table}
  \caption{
Initial data parameters for a $\chi=0.95$ highly spinning
  binary with mass ratio $q=2/3$. The two spins are given by $\vec S_i = (0,0,S_i)$ and the
two momenta are $\vec P_i = (P_i^r, P_i^t,0)$, where $i=1,2$.  The parameter $M$ is
the sum of the masses of the two black holes. Unlike for Bowen-York data, the
momenta and spins cannot be specified exactly. However, the mass
$M=m_1 + m_2$ is very close to the measured horizon mass $m_{H}$.
Quantities denoted by ``init'' were measured at
$t=0$, while quantities denoted by ``equi'' are measured
at $t=200$.  
$m^{H}_i$, $S_i$, $\chi_i$ are masses, spin angular momenta,
and dimensionless spins, respectively, of the two black holes.
The quantity $r_{H}$ is the polar coordinate radius of the
horizons.
Finally, $M_{\rm ADM}$ and $J_{\rm ADM}$ are the ADM masses and
spins. Also included are the attenuation and fisheye parameters
described in the text.
}\label{tab:id}
  \begin{ruledtabular}
    \begin{tabular}{llllll}
      Initial Data Quantities  \\
      \hline
      \\
      $P_1^r/M =  0.00101 $ &  $P_2^t/M = -0.097945$ \\
      $P_2^r/M = -0.00100 $ &  $P_1^t/M =  0.098958$ \\
      $m_1/M = 0.39860 $ & $m_2/M = 0.60140$ \\
      $S_1/M^2 = 0.15094 $ & $S_2/M^2 = -0.34359$ \\ 
      \\
      $J_{\rm ADM}/M^2 = 0.74449$     & $M_{\rm ADM}/M = 0.98873$\\
      $m_1^{H\ \rm init}/M = 0.39846$ & $m_2^{H\ \rm init}/M = 0.60019$ \\
      $S_1^{\rm init}/M^2 = 0.15090$  & $S_2^{\rm init}/M^2 = -0.34347$ \\
      $\chi_1^{\rm init} = 0.95042$   & $\chi_2^{\rm init} = -0.95346$ \\
      $r_1^{H\ \rm init}/M = 0.422 $  & $r_2^{H\ \rm init}/M = 0.420$\\
      \\
      \hline
      Relaxed Quantities \\
      \hline
      \\
      $m_1^{H\ \rm equi}/M = 0.3985\pm0.0001$ & $m_2^{H\ \rm equi}/M =  0.6002\pm0.0008$ \\
      $S_1^{\rm equi}/M^2 = 0.1518\pm0.0001$    & $S_2^{\rm equi}/M^2  = -0.3440\pm0.0004$ \\
      $\chi_1^{\rm equi} = 0.9503\pm0.0002$     & $\chi_2^{\rm equi}   = -0.9534\pm0.0006$ \\
      $r_1^{H\ \rm equi}/M = 0.173\pm0.001$     & $r_2^{H\ \rm equi}/M =  0.273\pm0.001$ \\
     \\
     \hline Additional Parameters\\
     \hline
      $r_{\rm min} = 0.01$ & $r_{\rm max} = 0.4$\\
    $A_{\rm FE 2} = 0.86$ &  $s_{\rm FE 2} = 1.5$\\
    $A_{\rm FE 1} = 0.936$ &  $s_{\rm FE 1} = 1.5$\\
      $s_{\rm far} = 10.0$\\

    \end{tabular}
  \end{ruledtabular}
\end{table}

\subsection{Evolution}\label{sec:evolution}

We evolve black hole binary initial data sets using the 
{\sc LazEv}~\cite{Zlochower:2005bj} implementation of the \MPA 
for the conformal and covariant formulation of the Z4 (CCZ4) system
(Ref.~\cite{Alic:2011gg}) which includes stronger damping of
the constraint violations than the standard BSSNOK~\cite{Nakamura87, Shibata95,
Baumgarte99} system.
For the run presented here, we use
centered, eighth-order accurate finite differencing in
space~\cite{Lousto:2007rj} and a fourth-order Runge-Kutta time
integrator. 
Our code
uses the {\sc Cactus}/{\sc EinsteinToolkit}~\cite{cactus_web,
einsteintoolkit} infrastructure.  We use the {\sc Carpet} mesh 
refinement driver to provide a ``moving boxes'' style of mesh refinement
\cite{Schnetter-etal-03b}.  Fifth-order Kreiss-Oliger dissipation is added to
evolved variables with dissipation coefficient $\epsilon=0.1$.
For the CCZ4 damping parameters, we chose
$\kappa_1 = 0.21$, $\kappa_2=0$, and $\kappa_3=0$
(see~\cite{Alic:2011gg}).

We locate the apparent horizons using the {\sc AHFinderDirect}
code~\cite{Thornburg2003:AH-finding} and measure the horizon spins
using the isolated horizon algorithm~\cite{Dreyer02a}.
We calculate the radiation scalar $\psi_4$ using the Antenna
thorn~\cite{Campanelli:2005ia, Baker:2001sf}.
We then extrapolate the waveform to
an infinite observer location using
the perturbative formulas given in Ref.~\cite{Nakano:2015pta}.

For the gauge equations, we use~\cite{Alcubierre02a,
Campanelli:2005dd, vanMeter:2006vi}
\begin{subequations}
    \label{eq:gauge}
      \begin{align}
           (\partial_t - \beta^i \partial_i) \alpha &= - 2 \alpha^2 K \; , \\
            \partial_t \beta^a &= \frac{3}{4} \tilde{\Gamma}^a - \eta
        \beta^a \; .
          \end{align}
\end{subequations}
Note that the lapse is not evolved with the standard 1+log form. Here
we multiply the rhs of the lapse equation by an additional factor of
$\alpha$. This has the effect of increasing the equilibrium
(coordinate) size of the horizons. For the initial values of shift, we
chose $\beta^i(t=0) =0$, while for the initial values of the lapse, we
chose an ad-hoc function
$\alpha(t=0) = \tilde \psi^{-2}$, where
$\tilde \psi = 1 + {\cal M}/(2 r_1) + {\cal M}/(2 r_2)$ and
$r_i$ is the coordinate distance to black hole $i$. For the function $\eta$,
we chose
\begin{equation}
  \eta(\vec r) = (\eta_c - \eta_o) \exp(-(r/\eta_s)^4) + \eta_o,
\end{equation}
where
$\eta_c = 2.0/M$, $\eta_s = 40.0M$, and $\eta_o = 0.25/M$. With this
choice, $\eta$ is small in the outer zones. As shown in
Ref.~\cite{Schnetter:2010cz}, the magnitude of $\eta$ limits how large
the timestep can be with $dt_{\rm max} \propto 1/\eta$. Since this
limit is independent of spatial resolution, it is only significant in
the very coarse outer zones where the standard Courant-Friedrichs-Lewy
condition would otherwise lead to a large value for $dt_{\rm max}$.

The  grid structure consisted of 11 levels of refinement
with the finest mesh extending to $\pm0.3M$ (in all directions) from the centers of the
two black holes with a grid spacing
of $M/368.64$, while the coarsest level extended to $\pm400M$ (in all
directions) with a grid
spacing of$M/0.36$.  
We removed one level around the larger black hole after it relaxed.  
The total run required 868,222SUs in our local machine, {\it Blue Sky}
on 32 nodes until merger, and then 24 nodes afterwards in a
wall-time of 69 days and resolution labeled as N144.

\section{Results}\label{sec:results}

We performed a single simulation from a coordinate separation of $10M$
(proper separation of $13.8M$) through merger for an unequal-mass binary,
$q=2/3$ where the larger hole spin is anti-aligned and the smaller aligned
with the orbital angular momentum and both
have dimensionless magnitudes of 0.95.

\begin{figure}[h!]
  \includegraphics[angle=270,width=0.40\textwidth]{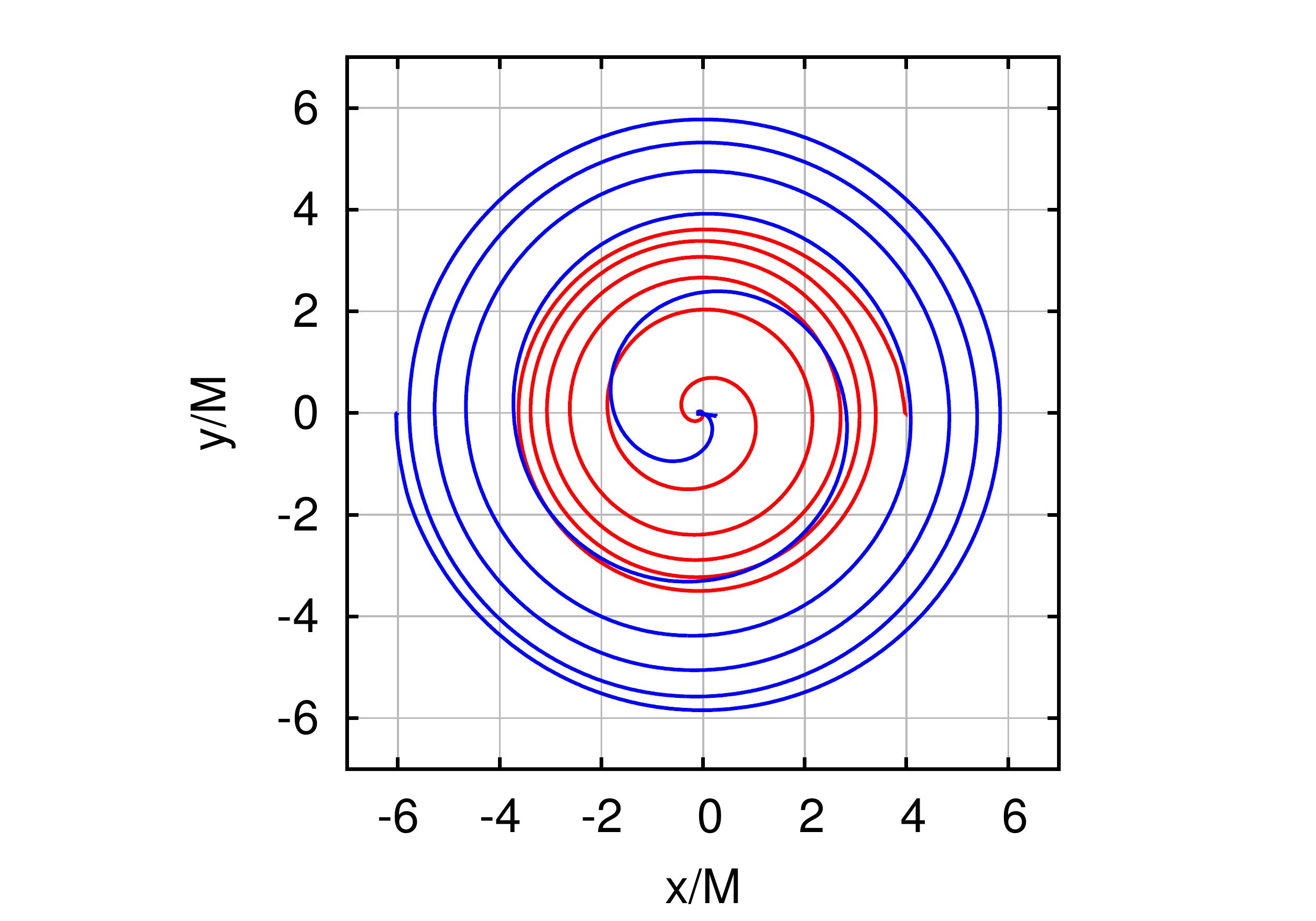}
  \includegraphics[angle=270,width=0.40\textwidth]{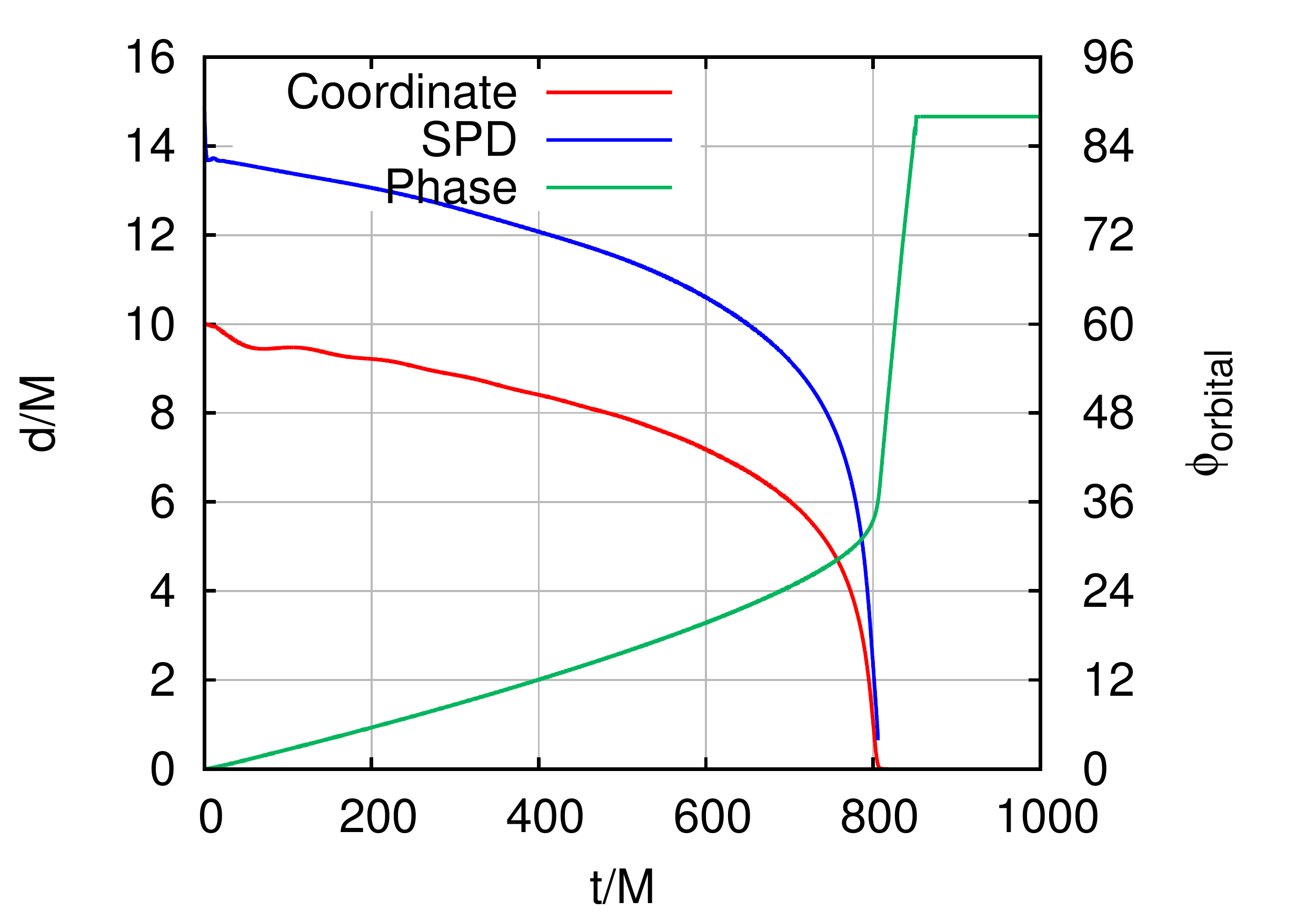}
  \caption{The trajectories of the two black holes, as well as the
    time dependence of the orbital separation (coordinate and simple proper distance) and phase.
\label{fig:tracks}}
\end{figure}

Figure \ref{fig:tracks} shows the tracks of the holes in the 
orbital (xy-) plane,
their relative  separation (both the coordinate separation and the simple proper distance along
the line joining the black holes), as well as the orbital phase.
To calculate the eccentricity, we fit a sinusoidal part and a secular part to the simple proper distance 
over a period of two orbits after the gauge settles (from $t=230M$ to $t=580M$).
The eccentricity is then $ e = |(D-D_{sec})/D| = 0.0013 $, 
where $D$ is the simple proper distance.

Note that we did not need to use an eccentricity reduction procedure like
\cite{Pfeiffer:2007yz, Buonanno:2010yk, Purrer:2012wy,
Buchman:2012dw} (although, this would be possible).  Rather, the initial data obtained using \hispid with
the parameters obtained by setting the radial momentum (pre-solve) and
post-solve net linear angular momentum to the values given by~\cite{Healy:2017zqj} is sufficient to obtain binaries with
eccentricity $\sim 0.001$. This shows that the improved procedure of \cite{Healy:2017zqj}
to provide quasicircular orbits, tested for lowers spins, also holds for the high spin binary
here considered.

\begin{figure*}[h!]
  \includegraphics[angle=270,width=0.32\textwidth]{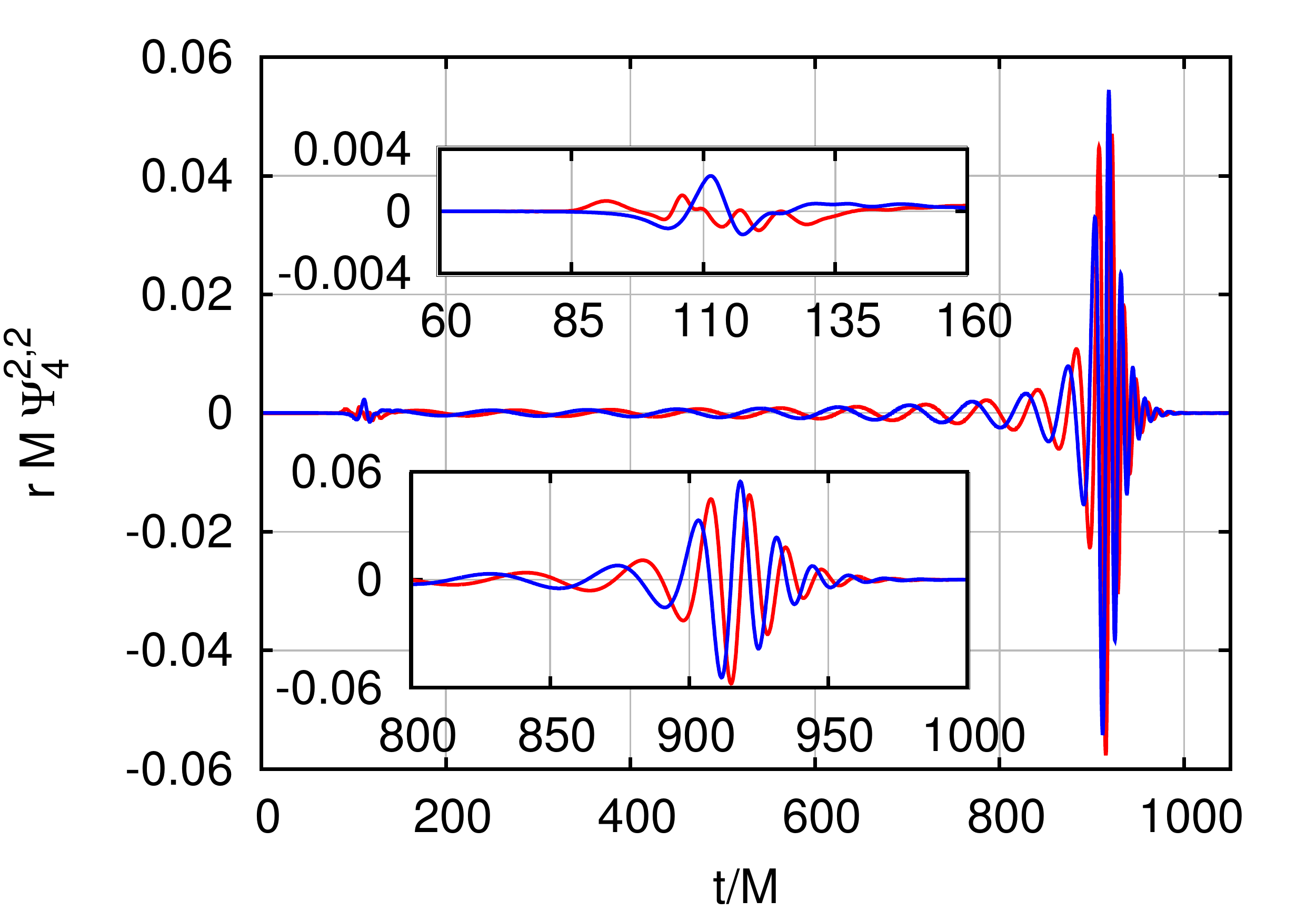}
  \includegraphics[angle=270,width=0.32\textwidth]{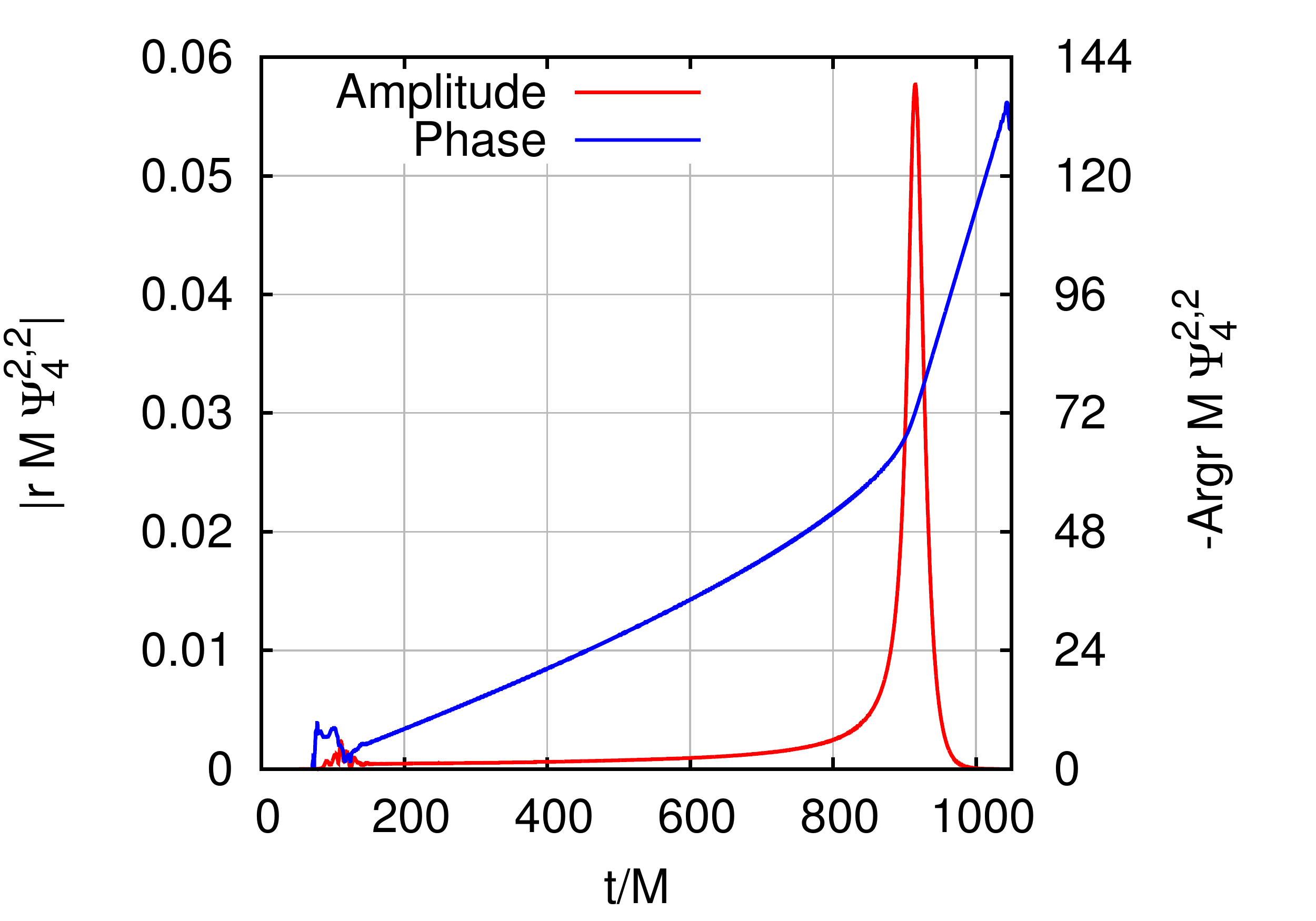}
  \includegraphics[angle=270,width=0.32\textwidth]{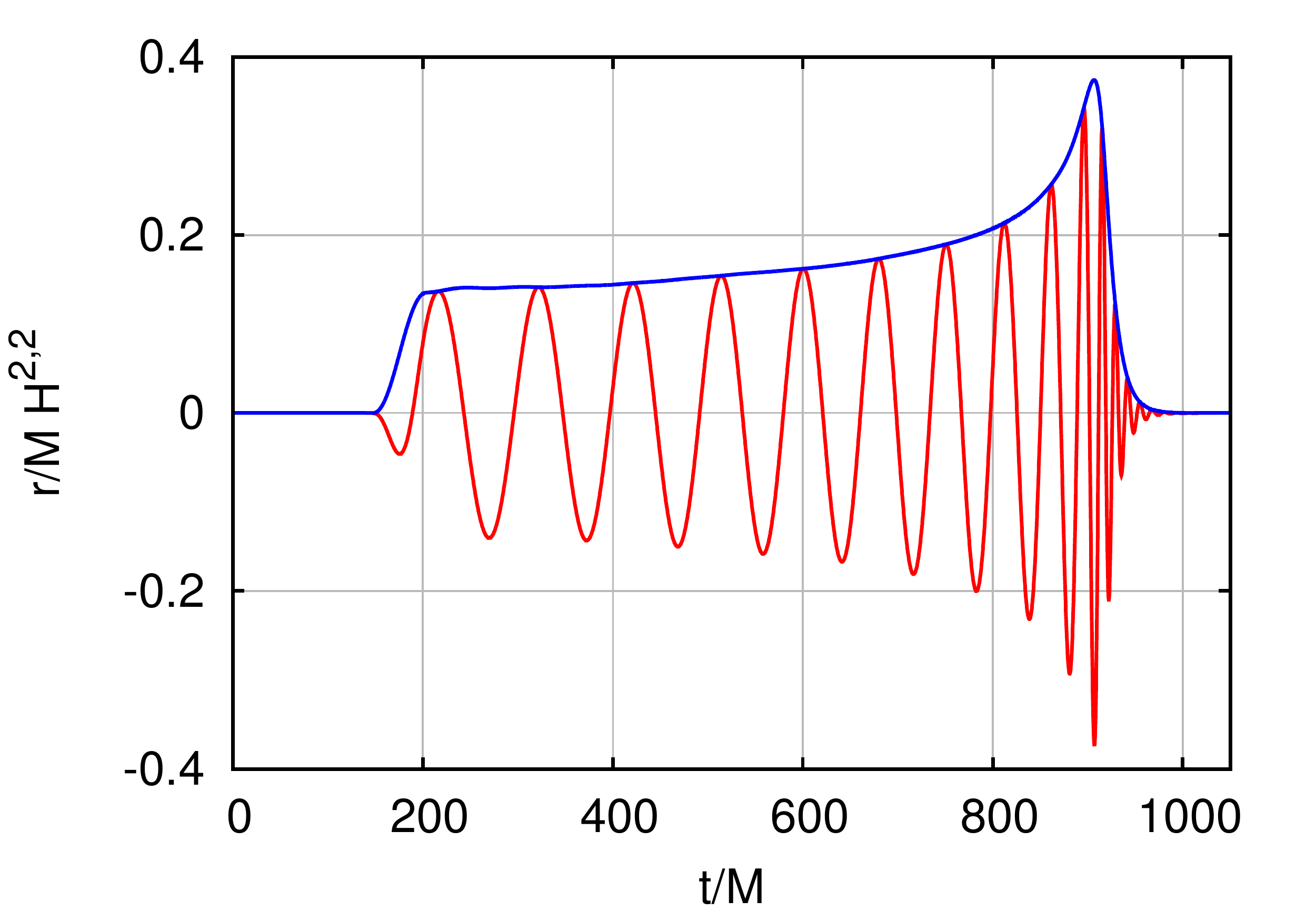}
  \caption{
    The (2,2) mode of $\psi_4$, its amplitude and reconstructed strain $h_{22}$
 as measured by an observer at location $r=102.6M$.  The 
strain in the right panel 
is extrapolated to infinite observer location using the analytic
perturbative extrapolation described in~\cite{Nakano:2015pta}.
\label{fig:waveforms}}
\end{figure*}

The waveform of the leading (2,2) mode is shown  
in Fig. \ref{fig:waveforms}.
We extract $\psi_4$ directly from the simulations,
and then compute the strain $h$ by double integration over time.
Note that at the relevant scale of the waveform, the initial
burst of radiation from our initial data is relatively small, almost
invisible.
This is in contrast for what is observed
in Bowen-York or other conformally flat initial data, where
for high spins, of the order of $0.9$, the initial burst can
have an amplitude comparable to that of the merger of the two
black holes and lead to serious contaminations of the evolution.
Besides, Bowen-York data cannot reach spin values of $0.95$ as
shown in this paper, since it is limited by spins below $0.93$
\cite{Cook:1989fb,Dain:2002ee,Lousto:2012es}.

\begin{figure*}[h!]
  \includegraphics[angle=270,width=0.32\textwidth]{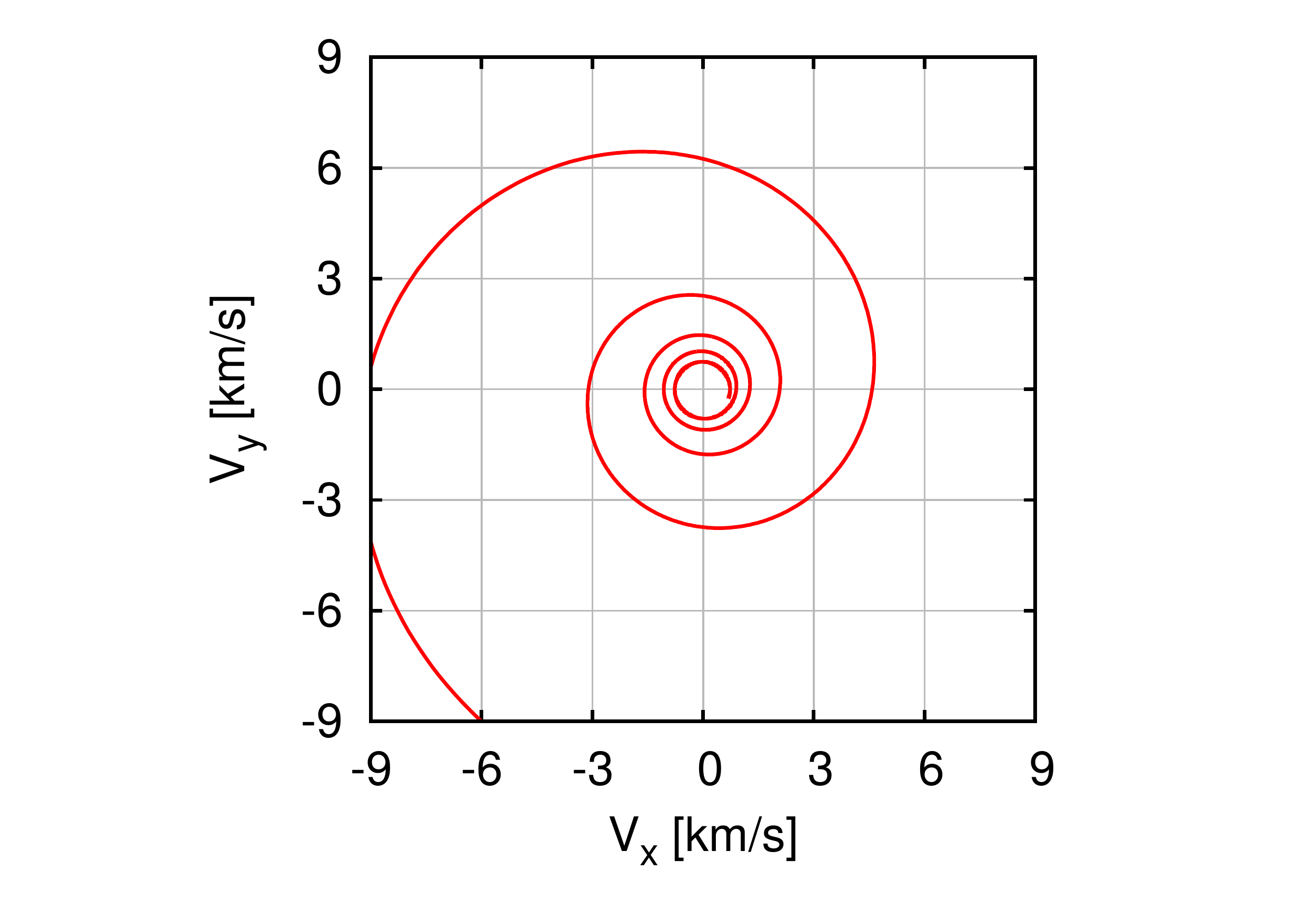}
  \includegraphics[angle=270,width=0.32\textwidth]{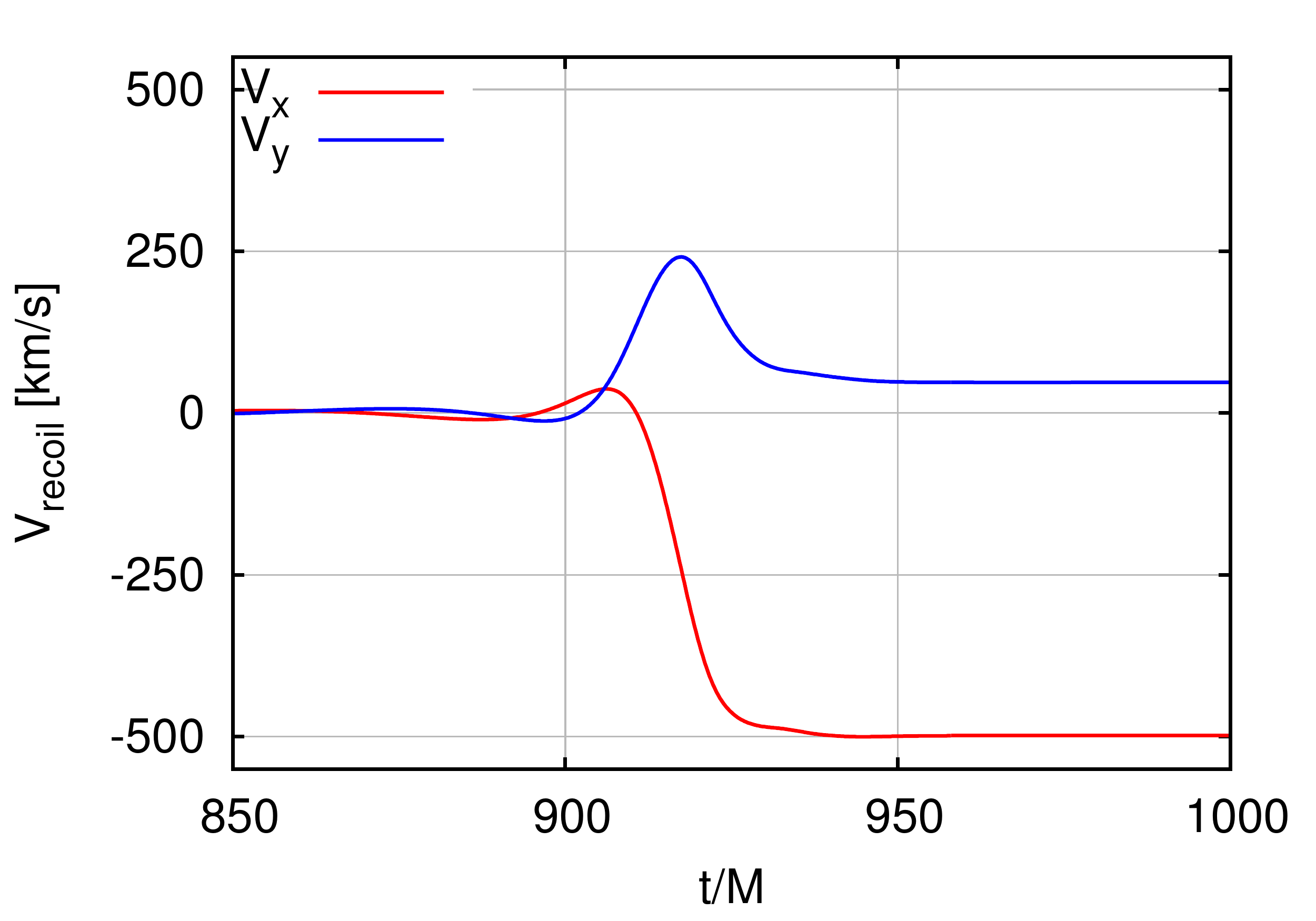}
  \includegraphics[angle=270,width=0.32\textwidth]{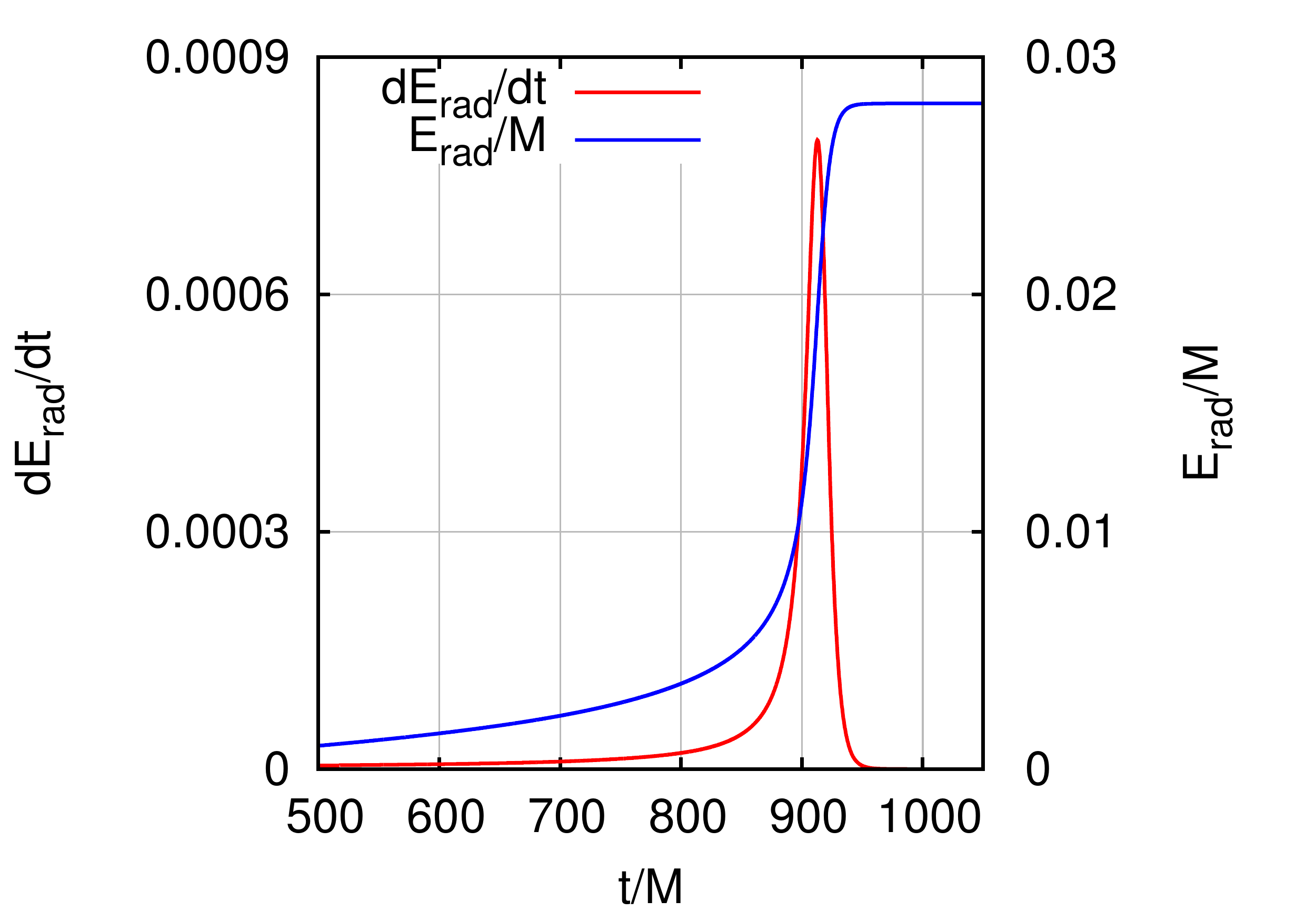}
  \caption{(left) The evolution in velocity space of the recoil
    vector during the inspiral. (middle) The cumulative recoil
    versus time. (right) The instantaneous radiated power and
    cumulative radiated energy versus time.
    All calculated at infinite observer location. \label{fig:recoil}}
\end{figure*}

From the waveforms we  compute the radiated energy and
radiated linear and angular momentum 
using the formulas given in \cite{Campanelli:1998jv, Lousto:2007mh}.
The recoil of the remnant is given by $-\delta \vec P/M_{\rm rem}$,
where $\delta \vec P$ is the radiated linear momentum and $M_{\rm
rem}$ is the mass of the remnant black hole.
Our results are summarized in Fig.~\ref{fig:recoil}.

\begin{figure*}
  \includegraphics[angle=270,width=0.32\textwidth]{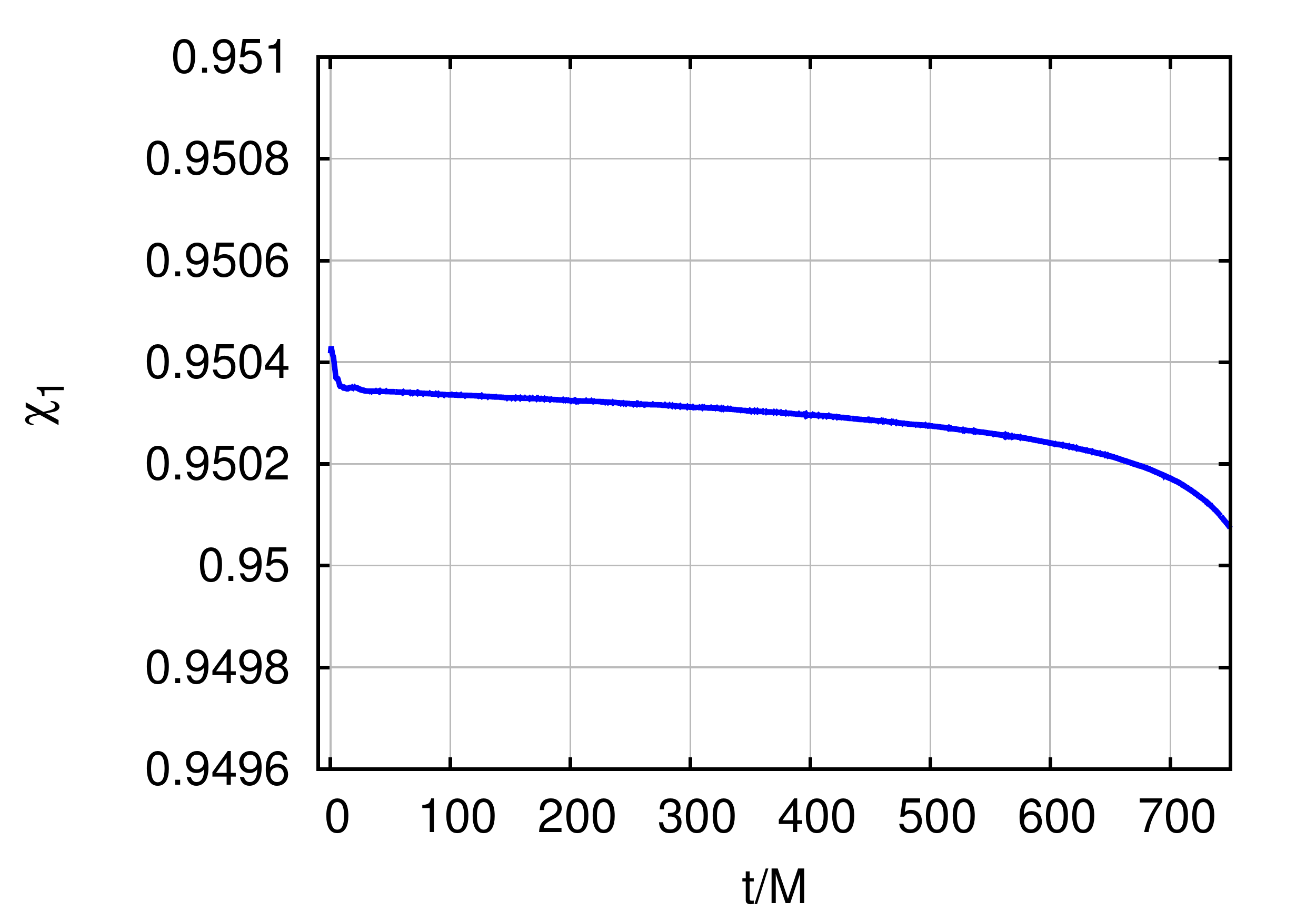}
  \includegraphics[angle=270,width=0.32\textwidth]{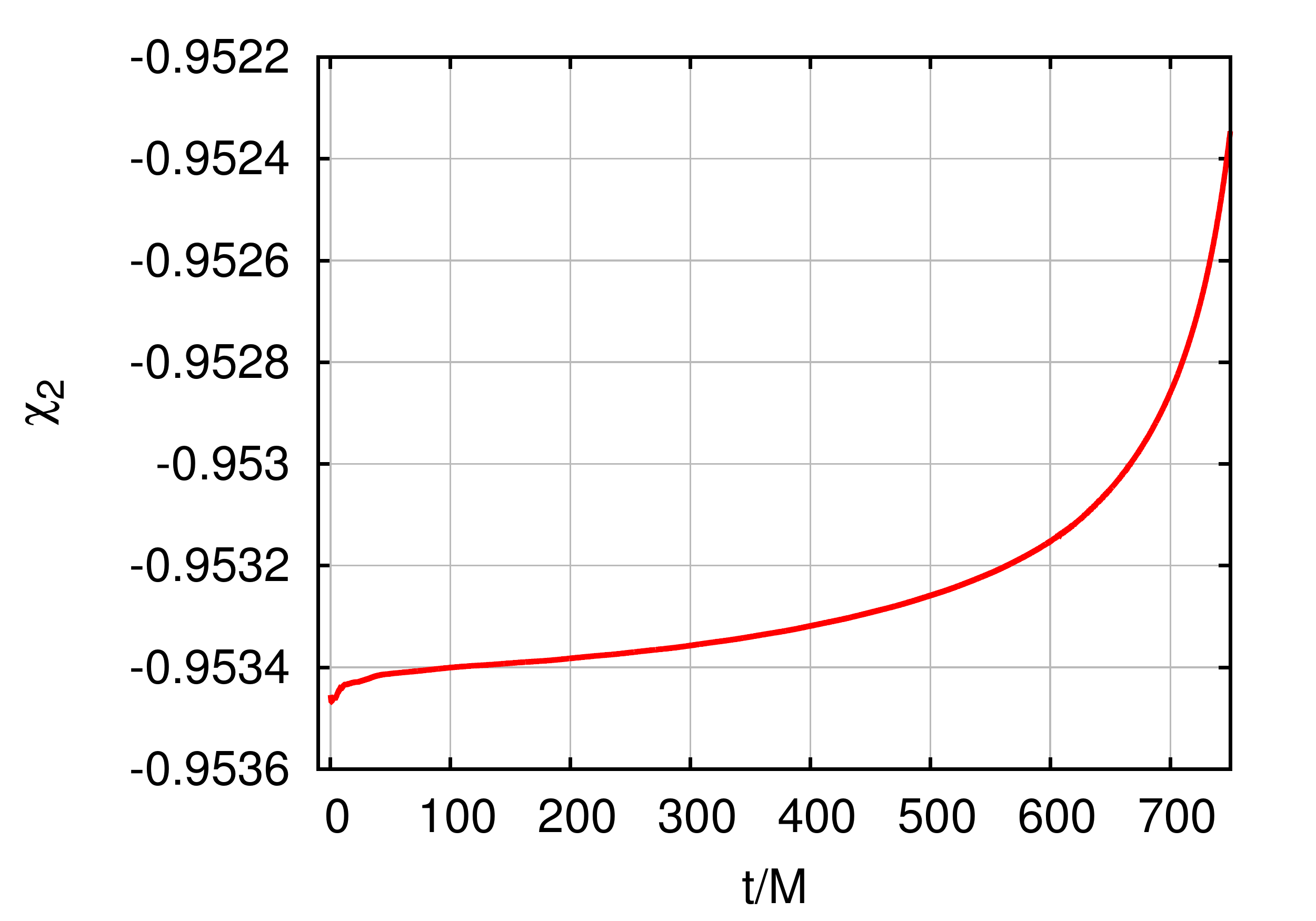}
  \includegraphics[angle=270,width=0.32\textwidth]{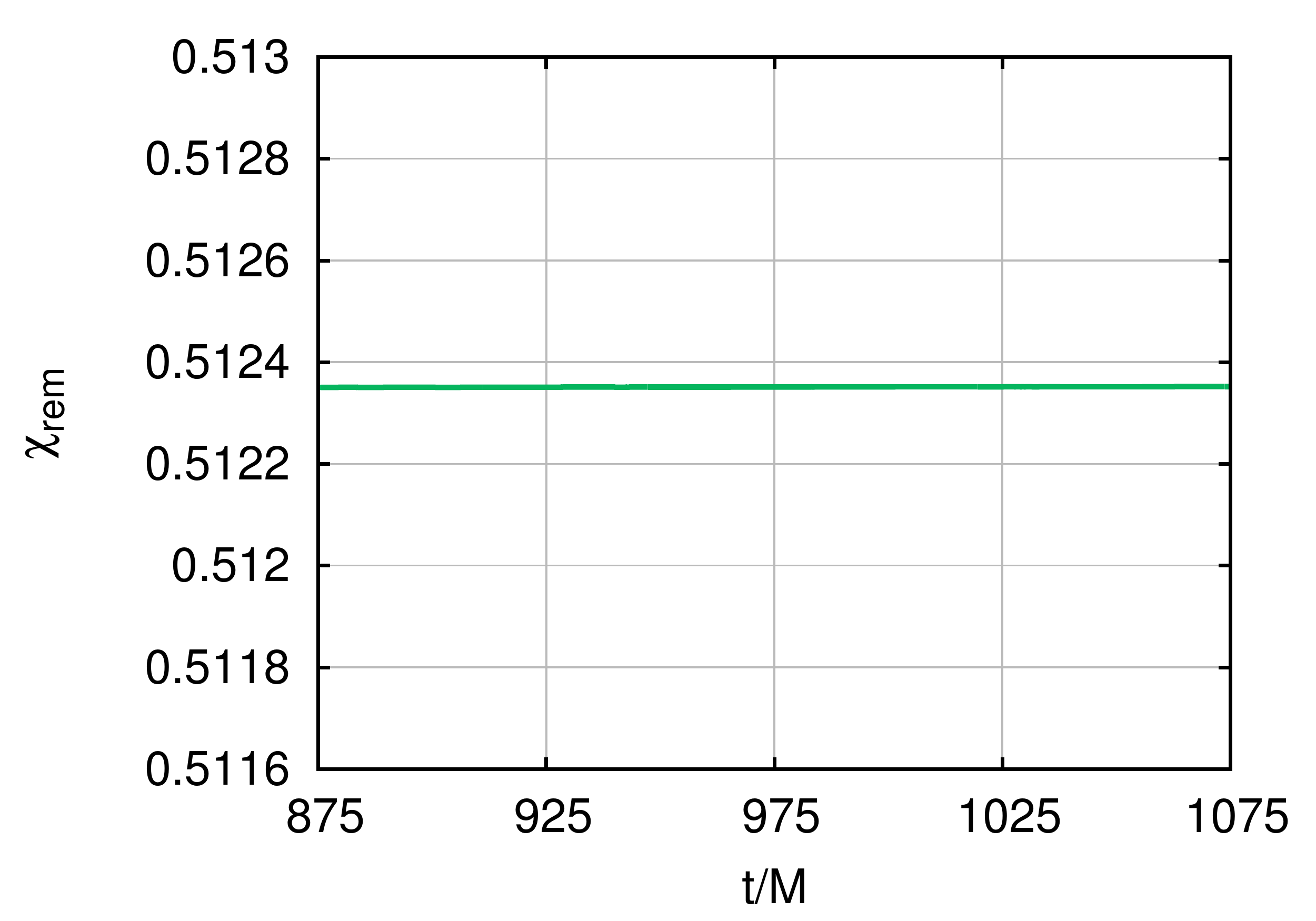}\\
  \includegraphics[angle=270,width=0.32\textwidth]{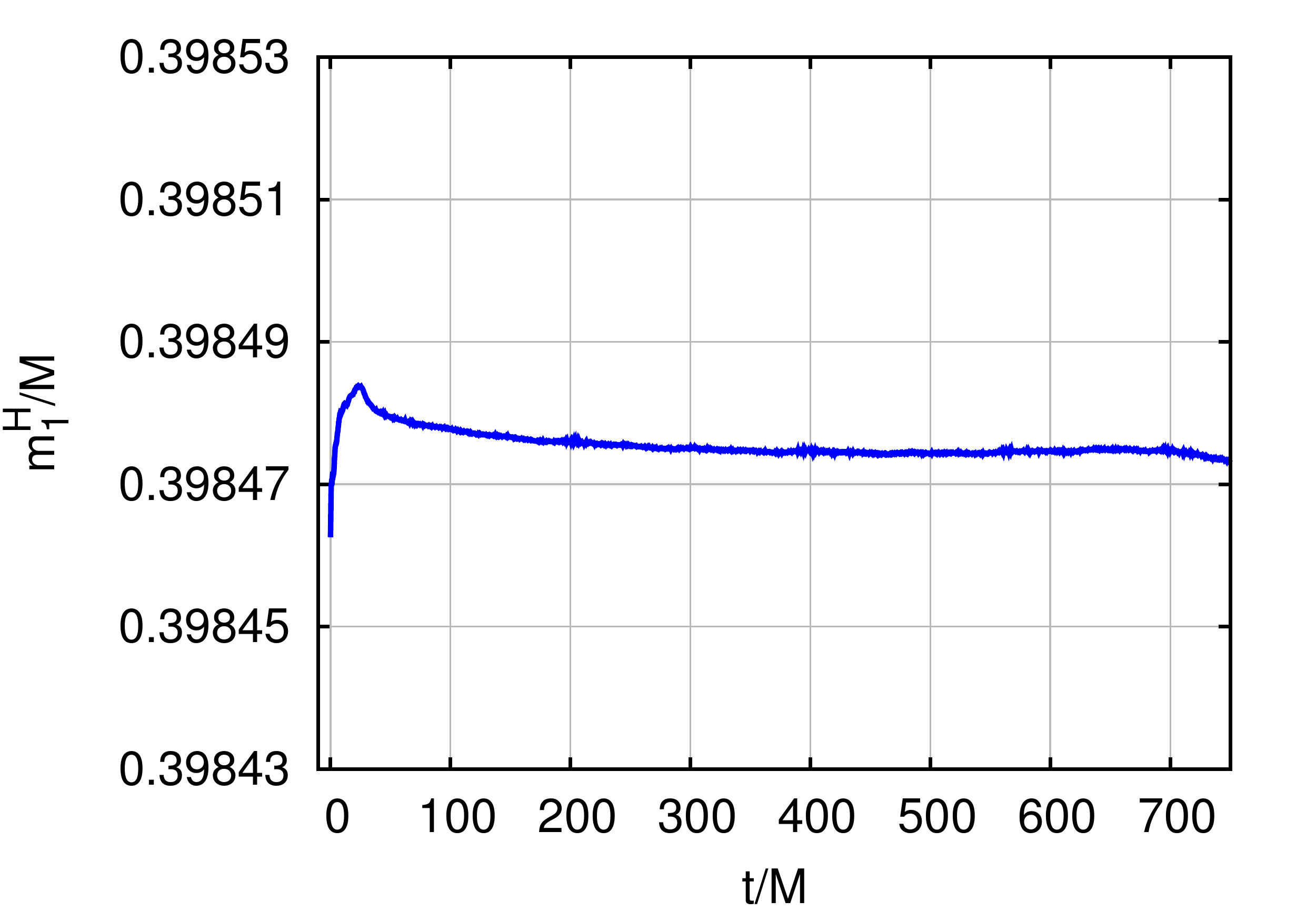}
  \includegraphics[angle=270,width=0.32\textwidth]{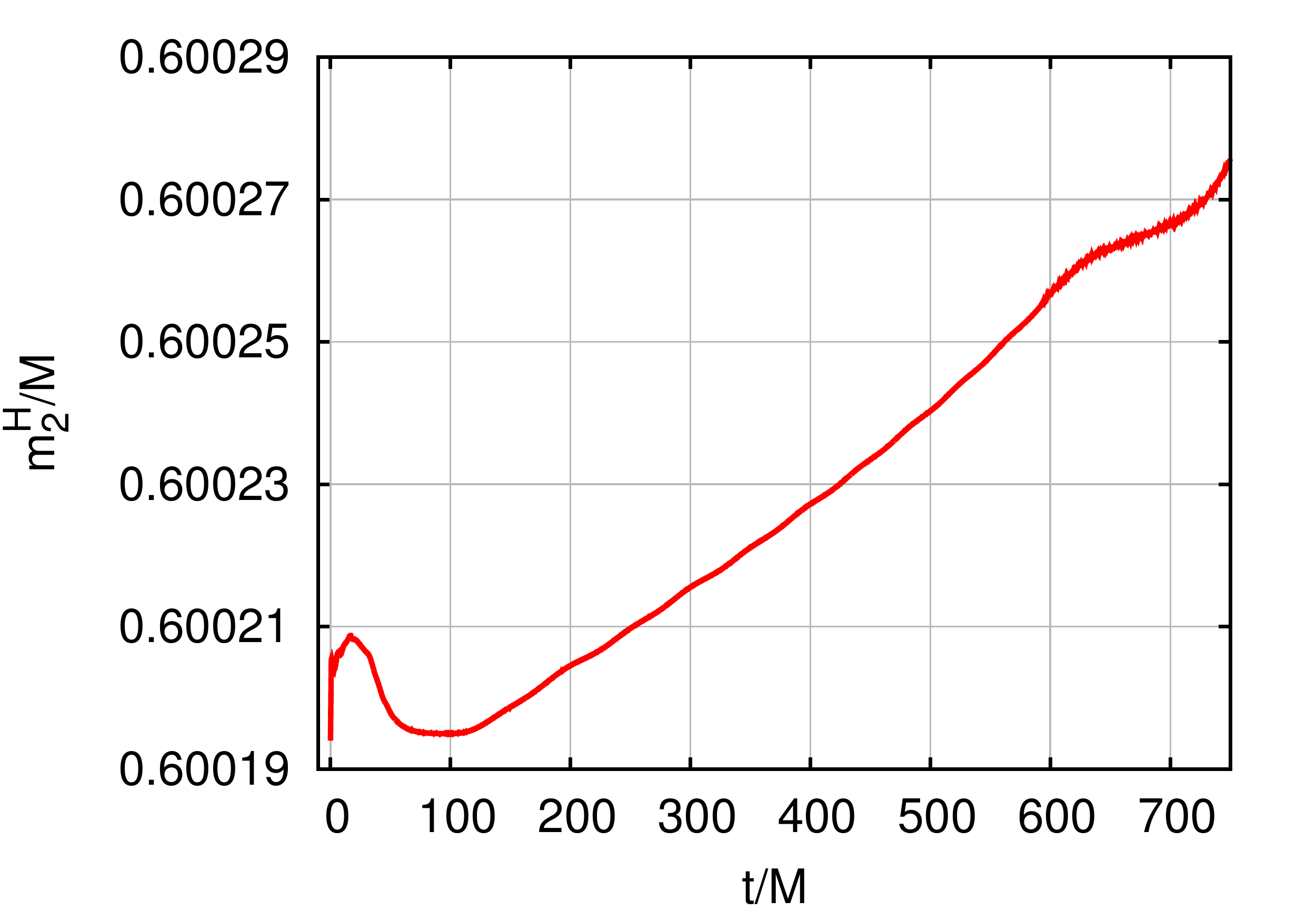}
  \includegraphics[angle=270,width=0.32\textwidth]{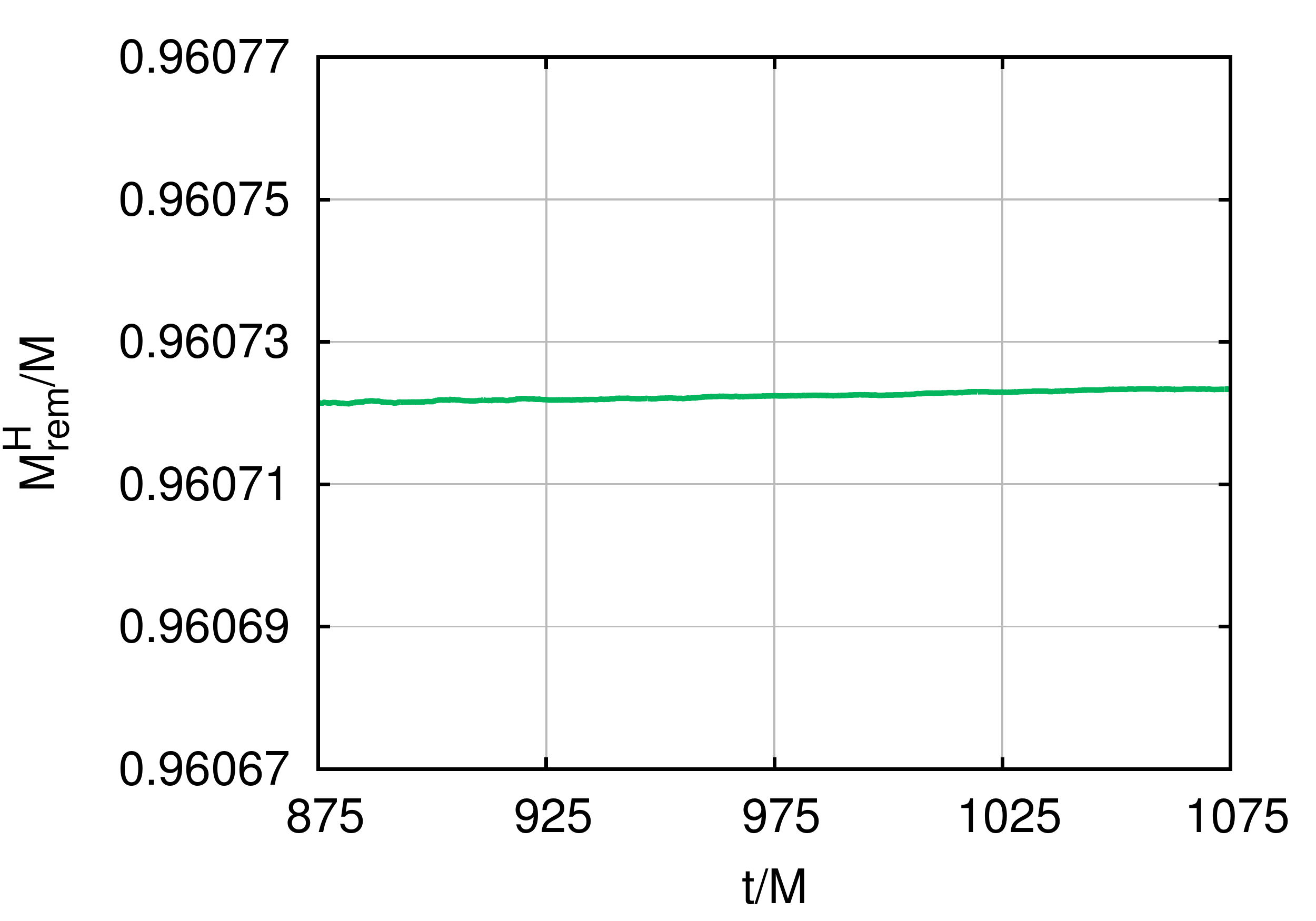}
  \caption{The dimensionless spin (top) and horizon (Christodoulou) mass 
  (bottom) for the two horizons in the binary and the final remnant black 
hole.  The $y$-scale for the mass and spins is set by black hole 2 and is kept the same for the other two black holes. }\label{fig:hor}
\end{figure*}

\begin{figure*}
  \includegraphics[angle=270,width=0.32\textwidth]{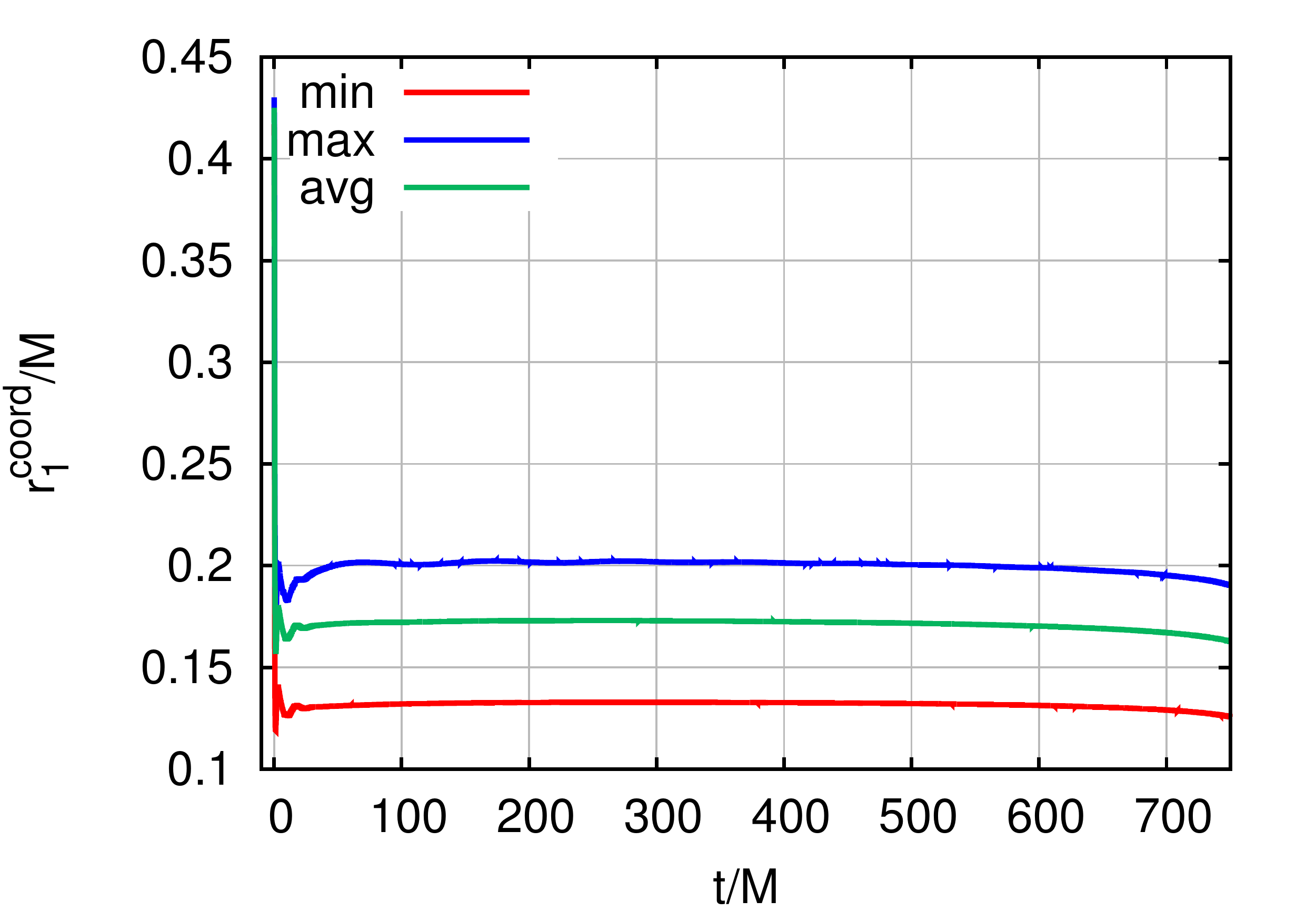}
  \includegraphics[angle=270,width=0.32\textwidth]{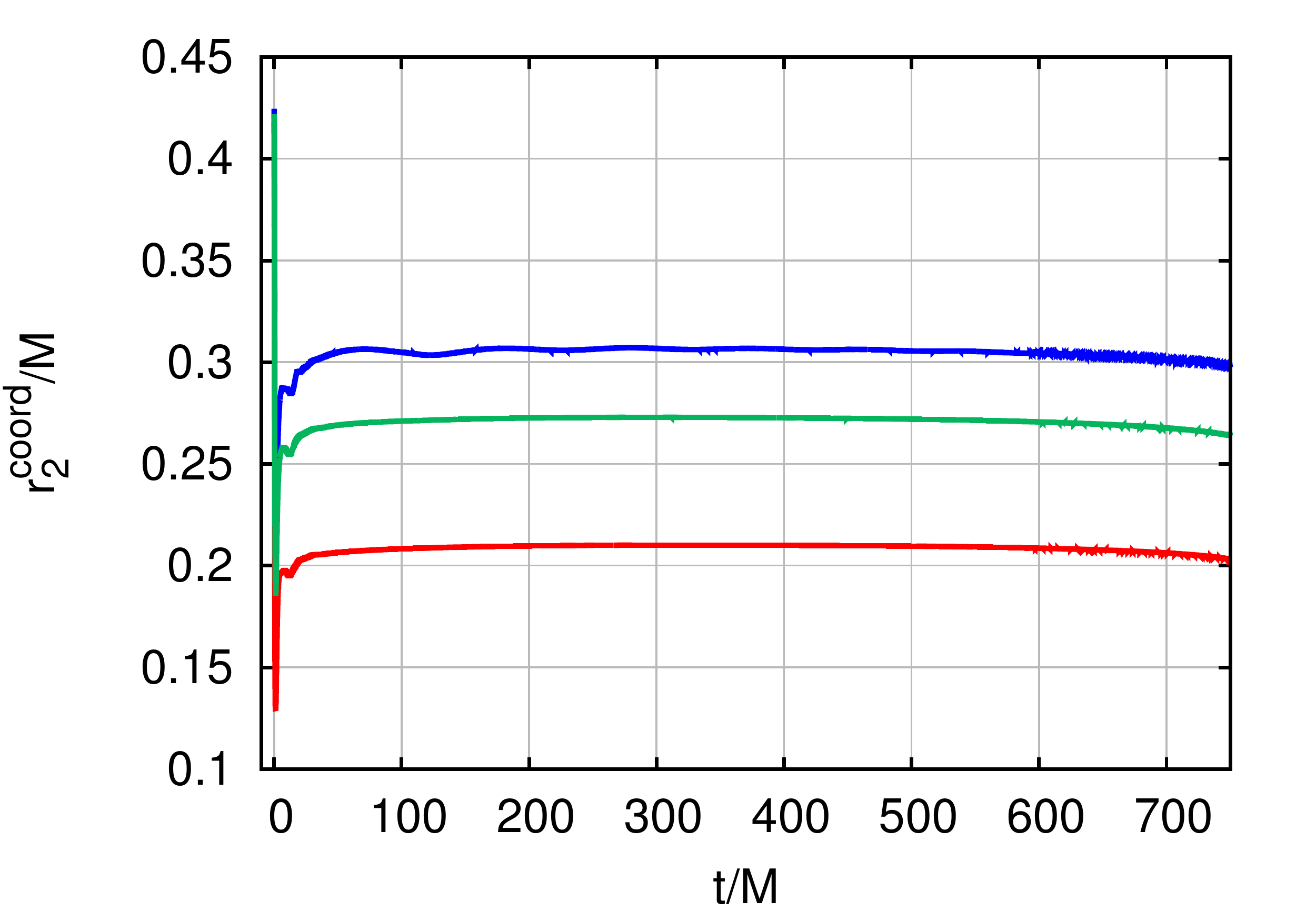}
  \includegraphics[angle=270,width=0.32\textwidth]{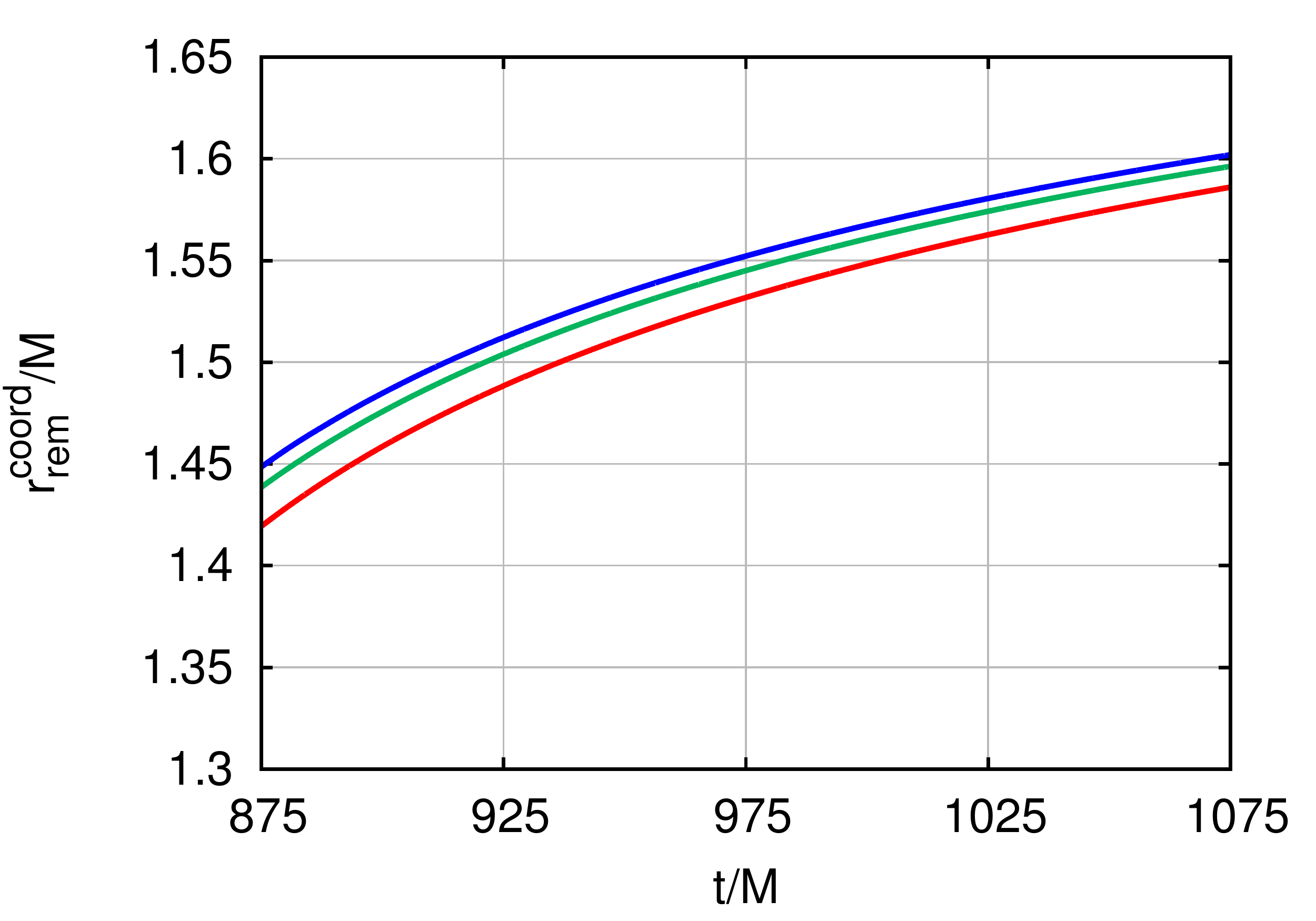}
  \caption{The coordinate radii (minimum, maximum and average) of 
the three horizons versus time for the full simulation. 
Note that there is an extremely rapid evolution of the horizon size
    and shape during the first few $M$ of evolution.
  \label{fig:horizon_radii}}
\end{figure*}

\begin{table}
  \caption{ 
Remnant quantities and comparison to fitting formulas.  
$M_{\rm rem}/M_{\rm equi}$ and $\chi_{\rm rem}$ are the final mass and spin 
of the remnant measured on the horizon.  
$V_{\rm recoil}$ and $L_{\rm peak}$ are the recoil velocity in km/s 
and the peak Luminosity in dimensionless units, measured at 
infinite observer location.  
$M_{\rm equi}\omega^{\rm peak}_{22}$ and $|r/M_{\rm equi} H^{\rm peak}_{22}|$ are the 
peak frequency and amplitude of the 22 mode of the strain.  The equilibrium
mass $M_{\rm equi} = m_1^{\rm equi} + m_2^{\rm equi} = 0.9987 \pm 0.0009$ is used for
:normalization.
}\label{tab:rem}
  \begin{ruledtabular}
    \begin{tabular}{llll}
    Quantity & Measured & Fit & \% difference \\
    \hline
    $M_{\rm rem}/M_{\rm equi}$           &  $0.9620\pm0.0009$           &  0.9620        & 0.00\% \\  
    $\chi_{\rm rem}$                 &  $0.512350\pm0.000002$       &  0.510031      & 0.45\% \\
    $V_{\rm recoil}$                 &  $500.10\pm0.49$             &  497.6         & 0.50\% \\
    $L_{\rm peak}(\times 10^{4})$     &  $7.952\pm0.0177$  & $7.840$  & 1.40\% \\
    $M_{\rm equi}\omega^{\rm peak}_{22}$ &  $0.3278\pm0.0019$           & 0.3309         & 0.90\% \\
    $|r/M_{\rm equi} H^{\rm peak}_{22}|$ &  $0.3749\pm0.0010$           & 0.3743         & 0.16\% \\
    \end{tabular}
  \end{ruledtabular}
\end{table}

\subsection{Diagnostics}\label{sec:diagnostic}

One of the most important diagnostics for a black-hole-binary simulation is the
degree to which the constraints are satisfied and to what degree the
horizon masses and spins are conserved. In Fig.~\ref{fig:hor}, we show the individual horizon mass and
dimensionless spin during the evolution, as well as the remnant mass
and spin post-merger. 
Due to our grid configuration, the smaller black hole was actually
better resolved. Consequently, the spin of the smaller black hole was
actually conserved to a better degree. The spin of the smaller black
hole decreased slowly for a net change of 0.0002, or $0.02\%$., the
larger black hole, on the other hand, showed a spin decrease (in
magnitude) of 0.001, or $0.1\%$. The smaller black hole's mass varied
by less than $0.005\%$, while the larger black hole's mass increased
by $0.013\%$. 
Note that prior to merger, the spins are within
$\pm0.003$ of $0.95$ and the masses change by less than 0.13\%, etc.

\begin{figure}
  \includegraphics[angle=270,width=0.44\textwidth]{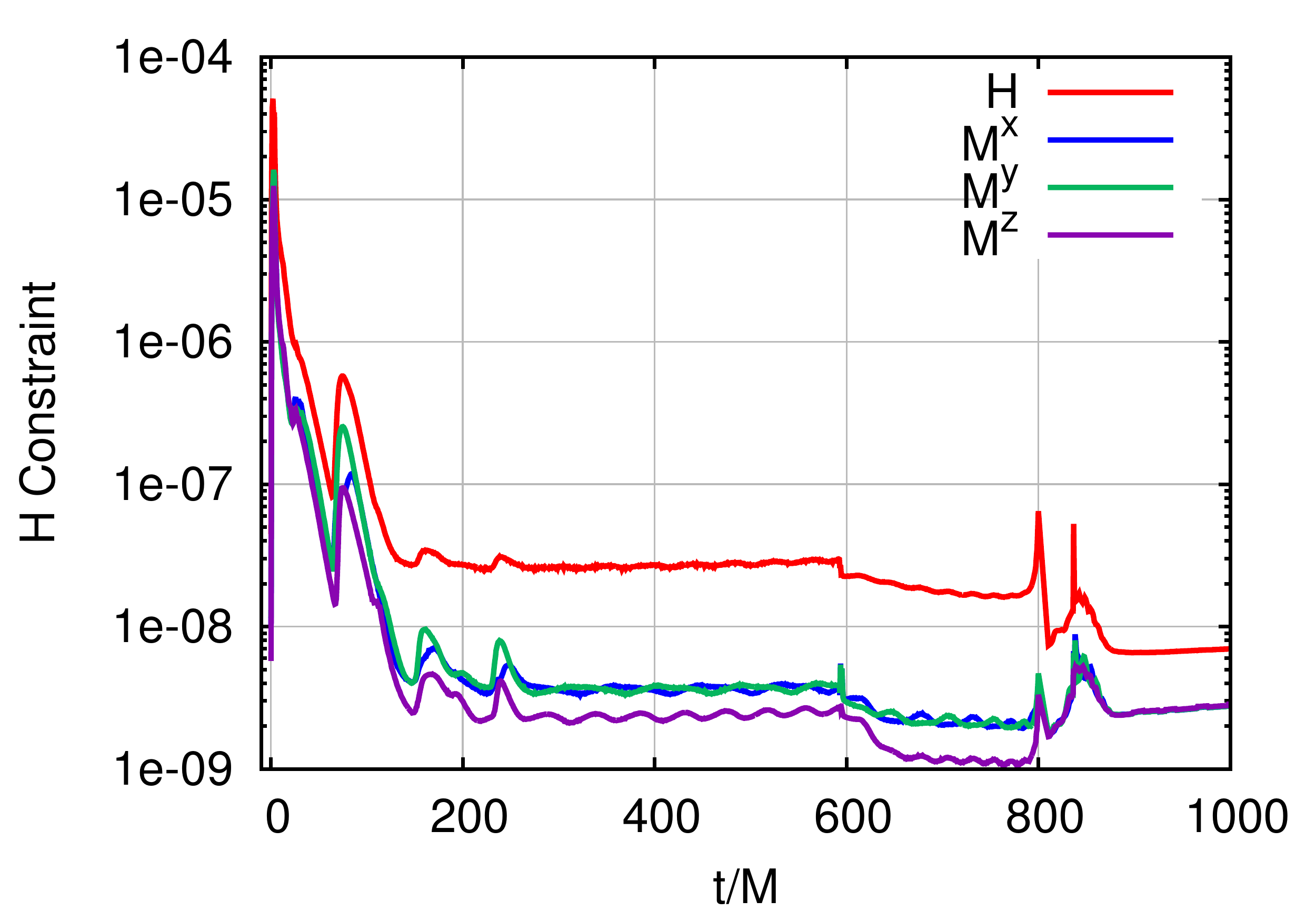}
  \caption{
$L^2$ norm of the Hamiltonian and momentum constraints
    versus time. Note the rapid growth during the first 2M of
    evolution. The CCZ4 damping 
parameters $\kappa_{1,2}$ managed to  suppress the constraint growths
    during the evolution down to merger and afterwards.}\label{fig:const}
\end{figure}

In Fig.~\ref{fig:const}, we show the $L^2$ norm of the Hamiltonian and
momentum constraints. Here the $L^2$ norm is over the region outside
the two horizons (or common horizon) and inside a sphere of radius
$30M$. Note how the constraints start small ($5\times10^{-9} - 5\times10^{-8}$) and quickly
increase to $10^{-5} - 10^{-4}$. This increase is due to unresolved features in
the initial data (i.e., the AMR grid cannot propagate high-frequency
data accurately). The constraints then damp to $5\times 10^{-8} - 5
\times 10^{-7}$ and remain roughly constant from then on.

One method which we found was useful for increasing the run speed was
to change the lapse condition. Rather than using the standard 1+log
lapse, we use a modified slicing closely related to harmonic slicing.
This alternative lapse keeps the horizons at a larger coordinate size
than 1+log. However, there is still a rapid decrease in the coordinate
size of the horizons at very early time. This rapid change in the gauge
(see
Fig.~\ref{fig:horizon_radii}) may be responsible for the initial jump
in the constraint violations seen in Fig.~\ref{fig:const}.

\section{Discussion}\label{sec:discussion}

In this paper we demonstrated that it is possible to evolve 
unequal-mass black-hole
binaries with spins well beyond the Bowen-York limit using the 
``moving puncture''
formalism, and to efficiently generate the initial data for such
binaries with low eccentricity without resorting to expensive
iterative eccentricity-reduction procedures.
This means that comparative studies of these challenging
evolutions by the two main methods to numerically solve the field equations of
general relativity field equations  (the generalized harmonic approach
used by SXS and various flavors of the ``moving punctures'' approach
used by many other groups) are now possible beyond the equal-mass case \cite{Ruchlin:2014zva, Zlochower:2017bbg}.  Independent
comparison, along the lines explored in \cite{Lovelace:2016uwp}, have
been very successful in demonstrating the accuracy and correctness of
moderate-spin black hole simulations.  These new techniques also open
the possibility of exploring a region of parameter space which is of
high interest for both astrophysical and gravitational wave studies. 

In addition, we computed the peak luminosity and frequency, which are key characteristic
features of the merger phase of the binary, and
have contributed to the remnant final black hole modeling
by evaluating the final mass, spin, and recoil of the merged black hole.
In particular we have computed the largest recoil velocity recorded 
of nonprecessing binaries, just above 500 km/s, as predicted by the 
extrapolation of the formulas given in \cite{Healy:2014yta}.
The agreement between the extrapolation of the the fitting formulae
and the measured values from this simulation, as shown in
Table \ref{tab:rem}, give us a measure of the expected accuracy
of this kind of simulations.


\acknowledgments 
The authors
gratefully acknowledge the National Science Foundation (NSF) for financial support from Grants No.\
PHY-1607520, No.\ PHY-1707946, No.\ ACI-1550436, No.\ AST-1516150,
No.\ ACI-1516125, No.\ PHY-1726215.  This work used the Extreme Science and Engineering
Discovery Environment (XSEDE) [allocation TG-PHY060027N], which is
supported by NSF grant No. ACI-1548562.
Computational resources were also provided by the NewHorizons and
BlueSky Clusters at the Rochester Institute of Technology, which were
supported by NSF grants No.\ PHY-0722703, No.\ DMS-0820923, No.\
AST-1028087, and No.\ PHY-1229173. 



\bibliographystyle{apsrev4-1}
\bibliography{../../../../Bibtex/references}

\begin{thebibliography}{56}%
\makeatletter
\providecommand \@ifxundefined [1]{%
 \@ifx{#1\undefined}
}%
\providecommand \@ifnum [1]{%
 \ifnum #1\expandafter \@firstoftwo
 \else \expandafter \@secondoftwo
 \fi
}%
\providecommand \@ifx [1]{%
 \ifx #1\expandafter \@firstoftwo
 \else \expandafter \@secondoftwo
 \fi
}%
\providecommand \natexlab [1]{#1}%
\providecommand \enquote  [1]{``#1''}%
\providecommand \bibnamefont  [1]{#1}%
\providecommand \bibfnamefont [1]{#1}%
\providecommand \citenamefont [1]{#1}%
\providecommand \href@noop [0]{\@secondoftwo}%
\providecommand \href [0]{\begingroup \@sanitize@url \@href}%
\providecommand \@href[1]{\@@startlink{#1}\@@href}%
\providecommand \@@href[1]{\endgroup#1\@@endlink}%
\providecommand \@sanitize@url [0]{\catcode `\\12\catcode `\$12\catcode
  `\&12\catcode `\#12\catcode `\^12\catcode `\_12\catcode `\%12\relax}%
\providecommand \@@startlink[1]{}%
\providecommand \@@endlink[0]{}%
\providecommand \url  [0]{\begingroup\@sanitize@url \@url }%
\providecommand \@url [1]{\endgroup\@href {#1}{\urlprefix }}%
\providecommand \urlprefix  [0]{URL }%
\providecommand \Eprint [0]{\href }%
\providecommand \doibase [0]{http://dx.doi.org/}%
\providecommand \selectlanguage [0]{\@gobble}%
\providecommand \bibinfo  [0]{\@secondoftwo}%
\providecommand \bibfield  [0]{\@secondoftwo}%
\providecommand \translation [1]{[#1]}%
\providecommand \BibitemOpen [0]{}%
\providecommand \bibitemStop [0]{}%
\providecommand \bibitemNoStop [0]{.\EOS\space}%
\providecommand \EOS [0]{\spacefactor3000\relax}%
\providecommand \BibitemShut  [1]{\csname bibitem#1\endcsname}%
\let\auto@bib@innerbib\@empty
\bibitem [{\citenamefont {Pretorius}(2005)}]{Pretorius:2005gq}%
  \BibitemOpen
  \bibfield  {author} {\bibinfo {author} {\bibfnamefont {F.}~\bibnamefont
  {Pretorius}},\ }\href@noop {} {\bibfield  {journal} {\bibinfo  {journal}
  {Phys. Rev. Lett.}\ }\textbf {\bibinfo {volume} {95}},\ \bibinfo {pages}
  {121101} (\bibinfo {year} {2005})},\ \Eprint
  {http://arxiv.org/abs/gr-qc/0507014} {gr-qc/0507014} \BibitemShut {NoStop}%
\bibitem [{\citenamefont {Campanelli}\ \emph
  {et~al.}(2006{\natexlab{a}})\citenamefont {Campanelli}, \citenamefont
  {Lousto}, \citenamefont {Marronetti},\ and\ \citenamefont
  {Zlochower}}]{Campanelli:2005dd}%
  \BibitemOpen
  \bibfield  {author} {\bibinfo {author} {\bibfnamefont {M.}~\bibnamefont
  {Campanelli}}, \bibinfo {author} {\bibfnamefont {C.~O.}\ \bibnamefont
  {Lousto}}, \bibinfo {author} {\bibfnamefont {P.}~\bibnamefont {Marronetti}},
  \ and\ \bibinfo {author} {\bibfnamefont {Y.}~\bibnamefont {Zlochower}},\
  }\href@noop {} {\bibfield  {journal} {\bibinfo  {journal} {Phys. Rev. Lett.}\
  }\textbf {\bibinfo {volume} {96}},\ \bibinfo {pages} {111101} (\bibinfo
  {year} {2006}{\natexlab{a}})},\ \Eprint {http://arxiv.org/abs/gr-qc/0511048}
  {gr-qc/0511048} \BibitemShut {NoStop}%
\bibitem [{\citenamefont {Baker}\ \emph {et~al.}(2006)\citenamefont {Baker},
  \citenamefont {Centrella}, \citenamefont {Choi}, \citenamefont {Koppitz},\
  and\ \citenamefont {van Meter}}]{Baker:2005vv}%
  \BibitemOpen
  \bibfield  {author} {\bibinfo {author} {\bibfnamefont {J.~G.}\ \bibnamefont
  {Baker}}, \bibinfo {author} {\bibfnamefont {J.}~\bibnamefont {Centrella}},
  \bibinfo {author} {\bibfnamefont {D.-I.}\ \bibnamefont {Choi}}, \bibinfo
  {author} {\bibfnamefont {M.}~\bibnamefont {Koppitz}}, \ and\ \bibinfo
  {author} {\bibfnamefont {J.}~\bibnamefont {van Meter}},\ }\href@noop {}
  {\bibfield  {journal} {\bibinfo  {journal} {Phys. Rev. Lett.}\ }\textbf
  {\bibinfo {volume} {96}},\ \bibinfo {pages} {111102} (\bibinfo {year}
  {2006})},\ \Eprint {http://arxiv.org/abs/gr-qc/0511103} {gr-qc/0511103}
  \BibitemShut {NoStop}%
\bibitem [{\citenamefont {Gonzalez}\ \emph {et~al.}(2009)\citenamefont
  {Gonzalez}, \citenamefont {Sperhake},\ and\ \citenamefont
  {Brugmann}}]{Gonzalez:2008bi}%
  \BibitemOpen
  \bibfield  {author} {\bibinfo {author} {\bibfnamefont {J.~A.}\ \bibnamefont
  {Gonzalez}}, \bibinfo {author} {\bibfnamefont {U.}~\bibnamefont {Sperhake}},
  \ and\ \bibinfo {author} {\bibfnamefont {B.}~\bibnamefont {Brugmann}},\
  }\href {\doibase 10.1103/PhysRevD.79.124006} {\bibfield  {journal} {\bibinfo
  {journal} {Phys. Rev.}\ }\textbf {\bibinfo {volume} {D79}},\ \bibinfo {pages}
  {124006} (\bibinfo {year} {2009})},\ \Eprint {http://arxiv.org/abs/0811.3952}
  {arXiv:0811.3952 [gr-qc]} \BibitemShut {NoStop}%
\bibitem [{\citenamefont {Lousto}\ \emph {et~al.}(2010)\citenamefont {Lousto},
  \citenamefont {Nakano}, \citenamefont {Zlochower},\ and\ \citenamefont
  {Campanelli}}]{Lousto:2010qx}%
  \BibitemOpen
  \bibfield  {author} {\bibinfo {author} {\bibfnamefont {C.~O.}\ \bibnamefont
  {Lousto}}, \bibinfo {author} {\bibfnamefont {H.}~\bibnamefont {Nakano}},
  \bibinfo {author} {\bibfnamefont {Y.}~\bibnamefont {Zlochower}}, \ and\
  \bibinfo {author} {\bibfnamefont {M.}~\bibnamefont {Campanelli}},\ }\href
  {\doibase 10.1103/PhysRevD.82.104057} {\bibfield  {journal} {\bibinfo
  {journal} {Phys. Rev.}\ }\textbf {\bibinfo {volume} {D82}},\ \bibinfo {pages}
  {104057} (\bibinfo {year} {2010})},\ \Eprint {http://arxiv.org/abs/1008.4360}
  {arXiv:1008.4360 [gr-qc]} \BibitemShut {NoStop}%
\bibitem [{\citenamefont {Lousto}\ and\ \citenamefont
  {Zlochower}(2011)}]{Lousto:2010ut}%
  \BibitemOpen
  \bibfield  {author} {\bibinfo {author} {\bibfnamefont {C.~O.}\ \bibnamefont
  {Lousto}}\ and\ \bibinfo {author} {\bibfnamefont {Y.}~\bibnamefont
  {Zlochower}},\ }\href {\doibase 10.1103/PhysRevLett.106.041101} {\bibfield
  {journal} {\bibinfo  {journal} {Phys. Rev. Lett.}\ }\textbf {\bibinfo
  {volume} {106}},\ \bibinfo {pages} {041101} (\bibinfo {year} {2011})},\
  \Eprint {http://arxiv.org/abs/1009.0292} {arXiv:1009.0292 [gr-qc]}
  \BibitemShut {NoStop}%
\bibitem [{\citenamefont {Sperhake}\ \emph {et~al.}(2011)\citenamefont
  {Sperhake}, \citenamefont {Cardoso}, \citenamefont {Ott}, \citenamefont
  {Schnetter},\ and\ \citenamefont {Witek}}]{Sperhake:2011ik}%
  \BibitemOpen
  \bibfield  {author} {\bibinfo {author} {\bibfnamefont {U.}~\bibnamefont
  {Sperhake}}, \bibinfo {author} {\bibfnamefont {V.}~\bibnamefont {Cardoso}},
  \bibinfo {author} {\bibfnamefont {C.~D.}\ \bibnamefont {Ott}}, \bibinfo
  {author} {\bibfnamefont {E.}~\bibnamefont {Schnetter}}, \ and\ \bibinfo
  {author} {\bibfnamefont {H.}~\bibnamefont {Witek}},\ }\href {\doibase
  10.1103/PhysRevD.84.084038} {\bibfield  {journal} {\bibinfo  {journal} {Phys.
  Rev.}\ }\textbf {\bibinfo {volume} {D84}},\ \bibinfo {pages} {084038}
  (\bibinfo {year} {2011})},\ \Eprint {http://arxiv.org/abs/1105.5391}
  {arXiv:1105.5391 [gr-qc]} \BibitemShut {NoStop}%
\bibitem [{\citenamefont {Chu}\ \emph {et~al.}(2016)\citenamefont {Chu},
  \citenamefont {Fong}, \citenamefont {Kumar}, \citenamefont {Pfeiffer},
  \citenamefont {Boyle}, \citenamefont {Hemberger}, \citenamefont {Kidder},
  \citenamefont {Scheel},\ and\ \citenamefont {Szilagyi}}]{Chu:2015kft}%
  \BibitemOpen
  \bibfield  {author} {\bibinfo {author} {\bibfnamefont {T.}~\bibnamefont
  {Chu}}, \bibinfo {author} {\bibfnamefont {H.}~\bibnamefont {Fong}}, \bibinfo
  {author} {\bibfnamefont {P.}~\bibnamefont {Kumar}}, \bibinfo {author}
  {\bibfnamefont {H.~P.}\ \bibnamefont {Pfeiffer}}, \bibinfo {author}
  {\bibfnamefont {M.}~\bibnamefont {Boyle}}, \bibinfo {author} {\bibfnamefont
  {D.~A.}\ \bibnamefont {Hemberger}}, \bibinfo {author} {\bibfnamefont {L.~E.}\
  \bibnamefont {Kidder}}, \bibinfo {author} {\bibfnamefont {M.~A.}\
  \bibnamefont {Scheel}}, \ and\ \bibinfo {author} {\bibfnamefont
  {B.}~\bibnamefont {Szilagyi}},\ }\href {\doibase
  10.1088/0264-9381/33/16/165001} {\bibfield  {journal} {\bibinfo  {journal}
  {Class. Quant. Grav.}\ }\textbf {\bibinfo {volume} {33}},\ \bibinfo {pages}
  {165001} (\bibinfo {year} {2016})},\ \Eprint
  {http://arxiv.org/abs/1512.06800} {arXiv:1512.06800 [gr-qc]} \BibitemShut
  {NoStop}%
\bibitem [{\citenamefont {Jani}\ \emph {et~al.}(2016)\citenamefont {Jani},
  \citenamefont {Healy}, \citenamefont {Clark}, \citenamefont {London},
  \citenamefont {Laguna},\ and\ \citenamefont {Shoemaker}}]{Jani:2016wkt}%
  \BibitemOpen
  \bibfield  {author} {\bibinfo {author} {\bibfnamefont {K.}~\bibnamefont
  {Jani}}, \bibinfo {author} {\bibfnamefont {J.}~\bibnamefont {Healy}},
  \bibinfo {author} {\bibfnamefont {J.~A.}\ \bibnamefont {Clark}}, \bibinfo
  {author} {\bibfnamefont {L.}~\bibnamefont {London}}, \bibinfo {author}
  {\bibfnamefont {P.}~\bibnamefont {Laguna}}, \ and\ \bibinfo {author}
  {\bibfnamefont {D.}~\bibnamefont {Shoemaker}},\ }\href {\doibase
  10.1088/0264-9381/33/20/204001} {\bibfield  {journal} {\bibinfo  {journal}
  {Class. Quant. Grav.}\ }\textbf {\bibinfo {volume} {33}},\ \bibinfo {pages}
  {204001} (\bibinfo {year} {2016})},\ \Eprint
  {http://arxiv.org/abs/1605.03204} {arXiv:1605.03204 [gr-qc]} \BibitemShut
  {NoStop}%
\bibitem [{\citenamefont {Lovelace}\ \emph {et~al.}(2008)\citenamefont
  {Lovelace}, \citenamefont {Owen}, \citenamefont {Pfeiffer},\ and\
  \citenamefont {Chu}}]{Lovelace:2008tw}%
  \BibitemOpen
  \bibfield  {author} {\bibinfo {author} {\bibfnamefont {G.}~\bibnamefont
  {Lovelace}}, \bibinfo {author} {\bibfnamefont {R.}~\bibnamefont {Owen}},
  \bibinfo {author} {\bibfnamefont {H.~P.}\ \bibnamefont {Pfeiffer}}, \ and\
  \bibinfo {author} {\bibfnamefont {T.}~\bibnamefont {Chu}},\ }\href {\doibase
  10.1103/PhysRevD.78.084017} {\bibfield  {journal} {\bibinfo  {journal} {Phys.
  Rev.}\ }\textbf {\bibinfo {volume} {D78}},\ \bibinfo {pages} {084017}
  (\bibinfo {year} {2008})},\ \Eprint {http://arxiv.org/abs/0805.4192}
  {arXiv:0805.4192 [gr-qc]} \BibitemShut {NoStop}%
\bibitem [{Note1()}]{Note1}%
  \BibitemOpen
  \bibinfo {note} {{\protect \tt https://www.black-holes.org}}\BibitemShut
  {NoStop}%
\bibitem [{\citenamefont {Cook}\ and\ \citenamefont
  {York}(1990)}]{Cook:1989fb}%
  \BibitemOpen
  \bibfield  {author} {\bibinfo {author} {\bibfnamefont {G.~B.}\ \bibnamefont
  {Cook}}\ and\ \bibinfo {author} {\bibfnamefont {J.}~\bibnamefont {York},
  \bibfnamefont {James~W.}},\ }\href@noop {} {\bibfield  {journal} {\bibinfo
  {journal} {Phys. Rev.}\ }\textbf {\bibinfo {volume} {D41}},\ \bibinfo {pages}
  {1077} (\bibinfo {year} {1990})}\BibitemShut {NoStop}%
\bibitem [{\citenamefont {Bowen}\ and\ \citenamefont
  {York}(1980)}]{Bowen:1980yu}%
  \BibitemOpen
  \bibfield  {author} {\bibinfo {author} {\bibfnamefont {J.~M.}\ \bibnamefont
  {Bowen}}\ and\ \bibinfo {author} {\bibfnamefont {J.~W.}\ \bibnamefont {York},
  \bibfnamefont {Jr.}},\ }\href {\doibase 10.1103/PhysRevD.21.2047} {\bibfield
  {journal} {\bibinfo  {journal} {Phys. Rev.}\ }\textbf {\bibinfo {volume}
  {D21}},\ \bibinfo {pages} {2047} (\bibinfo {year} {1980})}\BibitemShut
  {NoStop}%
\bibitem [{\citenamefont {Dain}\ \emph {et~al.}(2002)\citenamefont {Dain},
  \citenamefont {Lousto},\ and\ \citenamefont {Takahashi}}]{Dain:2002ee}%
  \BibitemOpen
  \bibfield  {author} {\bibinfo {author} {\bibfnamefont {S.}~\bibnamefont
  {Dain}}, \bibinfo {author} {\bibfnamefont {C.~O.}\ \bibnamefont {Lousto}}, \
  and\ \bibinfo {author} {\bibfnamefont {R.}~\bibnamefont {Takahashi}},\ }\href
  {\doibase 10.1103/PhysRevD.65.104038} {\bibfield  {journal} {\bibinfo
  {journal} {Phys. Rev.}\ }\textbf {\bibinfo {volume} {D65}},\ \bibinfo {pages}
  {104038} (\bibinfo {year} {2002})},\ \Eprint
  {http://arxiv.org/abs/gr-qc/0201062} {arXiv:gr-qc/0201062} \BibitemShut
  {NoStop}%
\bibitem [{\citenamefont {Lousto}\ \emph {et~al.}(2012)\citenamefont {Lousto},
  \citenamefont {Nakano}, \citenamefont {Zlochower}, \citenamefont {Mundim},\
  and\ \citenamefont {Campanelli}}]{Lousto:2012es}%
  \BibitemOpen
  \bibfield  {author} {\bibinfo {author} {\bibfnamefont {C.~O.}\ \bibnamefont
  {Lousto}}, \bibinfo {author} {\bibfnamefont {H.}~\bibnamefont {Nakano}},
  \bibinfo {author} {\bibfnamefont {Y.}~\bibnamefont {Zlochower}}, \bibinfo
  {author} {\bibfnamefont {B.~C.}\ \bibnamefont {Mundim}}, \ and\ \bibinfo
  {author} {\bibfnamefont {M.}~\bibnamefont {Campanelli}},\ }\href@noop {}
  {\bibfield  {journal} {\bibinfo  {journal} {Phys. Rev.}\ }\textbf {\bibinfo
  {volume} {D85}},\ \bibinfo {pages} {124013} (\bibinfo {year} {2012})},\
  \Eprint {http://arxiv.org/abs/1203.3223} {arXiv:1203.3223 [gr-qc]}
  \BibitemShut {NoStop}%
\bibitem [{\citenamefont {Lovelace}\ \emph {et~al.}(2012)\citenamefont
  {Lovelace}, \citenamefont {Boyle}, \citenamefont {Scheel},\ and\
  \citenamefont {Szilagyi}}]{Lovelace:2011nu}%
  \BibitemOpen
  \bibfield  {author} {\bibinfo {author} {\bibfnamefont {G.}~\bibnamefont
  {Lovelace}}, \bibinfo {author} {\bibfnamefont {M.}~\bibnamefont {Boyle}},
  \bibinfo {author} {\bibfnamefont {M.~A.}\ \bibnamefont {Scheel}}, \ and\
  \bibinfo {author} {\bibfnamefont {B.}~\bibnamefont {Szilagyi}},\ }\href@noop
  {} {\bibfield  {journal} {\bibinfo  {journal} {Class. Quant. Grav.}\ }\textbf
  {\bibinfo {volume} {29}},\ \bibinfo {pages} {045003} (\bibinfo {year}
  {2012})},\ \Eprint {http://arxiv.org/abs/1110.2229} {arXiv:1110.2229 [gr-qc]}
  \BibitemShut {NoStop}%
\bibitem [{\citenamefont {Scheel}\ \emph {et~al.}(2015)\citenamefont {Scheel},
  \citenamefont {Giesler}, \citenamefont {Hemberger}, \citenamefont {Lovelace},
  \citenamefont {Kuper}, \citenamefont {Boyle}, \citenamefont {Szil{\'a}gyi},\
  and\ \citenamefont {Kidder}}]{Scheel:2014ina}%
  \BibitemOpen
  \bibfield  {author} {\bibinfo {author} {\bibfnamefont {M.~A.}\ \bibnamefont
  {Scheel}}, \bibinfo {author} {\bibfnamefont {M.}~\bibnamefont {Giesler}},
  \bibinfo {author} {\bibfnamefont {D.~A.}\ \bibnamefont {Hemberger}}, \bibinfo
  {author} {\bibfnamefont {G.}~\bibnamefont {Lovelace}}, \bibinfo {author}
  {\bibfnamefont {K.}~\bibnamefont {Kuper}}, \bibinfo {author} {\bibfnamefont
  {M.}~\bibnamefont {Boyle}}, \bibinfo {author} {\bibfnamefont
  {B.}~\bibnamefont {Szil{\'a}gyi}}, \ and\ \bibinfo {author} {\bibfnamefont
  {L.~E.}\ \bibnamefont {Kidder}},\ }\href {\doibase
  10.1088/0264-9381/32/10/105009} {\bibfield  {journal} {\bibinfo  {journal}
  {Class. Quant. Grav.}\ }\textbf {\bibinfo {volume} {32}},\ \bibinfo {pages}
  {105009} (\bibinfo {year} {2015})},\ \Eprint {http://arxiv.org/abs/1412.1803}
  {arXiv:1412.1803 [gr-qc]} \BibitemShut {NoStop}%
\bibitem [{\citenamefont {Ruchlin}\ \emph {et~al.}(2017)\citenamefont
  {Ruchlin}, \citenamefont {Healy}, \citenamefont {Lousto},\ and\ \citenamefont
  {Zlochower}}]{Ruchlin:2014zva}%
  \BibitemOpen
  \bibfield  {author} {\bibinfo {author} {\bibfnamefont {I.}~\bibnamefont
  {Ruchlin}}, \bibinfo {author} {\bibfnamefont {J.}~\bibnamefont {Healy}},
  \bibinfo {author} {\bibfnamefont {C.~O.}\ \bibnamefont {Lousto}}, \ and\
  \bibinfo {author} {\bibfnamefont {Y.}~\bibnamefont {Zlochower}},\ }\href
  {\doibase 10.1103/PhysRevD.95.024033} {\bibfield  {journal} {\bibinfo
  {journal} {Phys. Rev.}\ }\textbf {\bibinfo {volume} {D95}},\ \bibinfo {pages}
  {024033} (\bibinfo {year} {2017})},\ \Eprint {http://arxiv.org/abs/1410.8607}
  {arXiv:1410.8607 [gr-qc]} \BibitemShut {NoStop}%
\bibitem [{\citenamefont {Healy}\ \emph {et~al.}(2016)\citenamefont {Healy},
  \citenamefont {Ruchlin}, \citenamefont {Lousto},\ and\ \citenamefont
  {Zlochower}}]{Healy:2015mla}%
  \BibitemOpen
  \bibfield  {author} {\bibinfo {author} {\bibfnamefont {J.}~\bibnamefont
  {Healy}}, \bibinfo {author} {\bibfnamefont {I.}~\bibnamefont {Ruchlin}},
  \bibinfo {author} {\bibfnamefont {C.~O.}\ \bibnamefont {Lousto}}, \ and\
  \bibinfo {author} {\bibfnamefont {Y.}~\bibnamefont {Zlochower}},\ }\href
  {\doibase 10.1103/PhysRevD.94.104020} {\bibfield  {journal} {\bibinfo
  {journal} {Phys. Rev.}\ }\textbf {\bibinfo {volume} {D94}},\ \bibinfo {pages}
  {104020} (\bibinfo {year} {2016})},\ \Eprint
  {http://arxiv.org/abs/1506.06153} {arXiv:1506.06153 [gr-qc]} \BibitemShut
  {NoStop}%
\bibitem [{\citenamefont {Zlochower}\ \emph {et~al.}(2017)\citenamefont
  {Zlochower}, \citenamefont {Healy}, \citenamefont {Lousto},\ and\
  \citenamefont {Ruchlin}}]{Zlochower:2017bbg}%
  \BibitemOpen
  \bibfield  {author} {\bibinfo {author} {\bibfnamefont {Y.}~\bibnamefont
  {Zlochower}}, \bibinfo {author} {\bibfnamefont {J.}~\bibnamefont {Healy}},
  \bibinfo {author} {\bibfnamefont {C.~O.}\ \bibnamefont {Lousto}}, \ and\
  \bibinfo {author} {\bibfnamefont {I.}~\bibnamefont {Ruchlin}},\ }\href
  {\doibase 10.1103/PhysRevD.96.044002} {\bibfield  {journal} {\bibinfo
  {journal} {Phys. Rev.}\ }\textbf {\bibinfo {volume} {D96}},\ \bibinfo {pages}
  {044002} (\bibinfo {year} {2017})},\ \Eprint
  {http://arxiv.org/abs/1706.01980} {arXiv:1706.01980 [gr-qc]} \BibitemShut
  {NoStop}%
\bibitem [{\citenamefont {Lange}\ \emph {et~al.}(2017)\citenamefont {Lange}
  \emph {et~al.}}]{Lange:2017wki}%
  \BibitemOpen
  \bibfield  {author} {\bibinfo {author} {\bibfnamefont {J.}~\bibnamefont
  {Lange}} \emph {et~al.},\ }\href@noop {} {\  (\bibinfo {year} {2017})},\
  \bibinfo {note} {accepted to Phys. Rev. D},\ \Eprint
  {http://arxiv.org/abs/1705.09833} {arXiv:1705.09833 [gr-qc]} \BibitemShut
  {NoStop}%
\bibitem [{\citenamefont {Healy}\ \emph
  {et~al.}(2017{\natexlab{a}})\citenamefont {Healy}, \citenamefont {Lousto},
  \citenamefont {Zlochower},\ and\ \citenamefont {Campanelli}}]{Healy:2017psd}%
  \BibitemOpen
  \bibfield  {author} {\bibinfo {author} {\bibfnamefont {J.}~\bibnamefont
  {Healy}}, \bibinfo {author} {\bibfnamefont {C.~O.}\ \bibnamefont {Lousto}},
  \bibinfo {author} {\bibfnamefont {Y.}~\bibnamefont {Zlochower}}, \ and\
  \bibinfo {author} {\bibfnamefont {M.}~\bibnamefont {Campanelli}},\ }\href
  {\doibase 10.1088/1361-6382/aa91b1} {\bibfield  {journal} {\bibinfo
  {journal} {Class. Quant. Grav.}\ }\textbf {\bibinfo {volume} {34}},\ \bibinfo
  {pages} {224001} (\bibinfo {year} {2017}{\natexlab{a}})},\ \Eprint
  {http://arxiv.org/abs/1703.03423} {arXiv:1703.03423 [gr-qc]} \BibitemShut
  {NoStop}%
\bibitem [{\citenamefont {Campanelli}\ \emph
  {et~al.}(2006{\natexlab{b}})\citenamefont {Campanelli}, \citenamefont
  {Lousto},\ and\ \citenamefont {Zlochower}}]{Campanelli:2006uy}%
  \BibitemOpen
  \bibfield  {author} {\bibinfo {author} {\bibfnamefont {M.}~\bibnamefont
  {Campanelli}}, \bibinfo {author} {\bibfnamefont {C.~O.}\ \bibnamefont
  {Lousto}}, \ and\ \bibinfo {author} {\bibfnamefont {Y.}~\bibnamefont
  {Zlochower}},\ }\href@noop {} {\bibfield  {journal} {\bibinfo  {journal}
  {Phys. Rev.}\ }\textbf {\bibinfo {volume} {D74}},\ \bibinfo {pages}
  {041501(R)} (\bibinfo {year} {2006}{\natexlab{b}})},\ \Eprint
  {http://arxiv.org/abs/gr-qc/0604012} {gr-qc/0604012} \BibitemShut {NoStop}%
\bibitem [{\citenamefont {Healy}\ \emph {et~al.}(2014)\citenamefont {Healy},
  \citenamefont {Lousto},\ and\ \citenamefont {Zlochower}}]{Healy:2014yta}%
  \BibitemOpen
  \bibfield  {author} {\bibinfo {author} {\bibfnamefont {J.}~\bibnamefont
  {Healy}}, \bibinfo {author} {\bibfnamefont {C.~O.}\ \bibnamefont {Lousto}}, \
  and\ \bibinfo {author} {\bibfnamefont {Y.}~\bibnamefont {Zlochower}},\ }\href
  {\doibase 10.1103/PhysRevD.90.104004} {\bibfield  {journal} {\bibinfo
  {journal} {Phys. Rev.}\ }\textbf {\bibinfo {volume} {D90}},\ \bibinfo {pages}
  {104004} (\bibinfo {year} {2014})},\ \Eprint {http://arxiv.org/abs/1406.7295}
  {arXiv:1406.7295 [gr-qc]} \BibitemShut {NoStop}%
\bibitem [{\citenamefont {Abbott}\ \emph {et~al.}(2016)\citenamefont {Abbott}
  \emph {et~al.}}]{Abbott:2016apu}%
  \BibitemOpen
  \bibfield  {author} {\bibinfo {author} {\bibfnamefont {B.~P.}\ \bibnamefont
  {Abbott}} \emph {et~al.} (\bibinfo {collaboration} {Virgo, LIGO
  Scientific}),\ }\href {\doibase 10.1103/PhysRevD.94.064035} {\bibfield
  {journal} {\bibinfo  {journal} {Phys. Rev.}\ }\textbf {\bibinfo {volume}
  {D94}},\ \bibinfo {pages} {064035} (\bibinfo {year} {2016})},\ \Eprint
  {http://arxiv.org/abs/1606.01262} {arXiv:1606.01262 [gr-qc]} \BibitemShut
  {NoStop}%
\bibitem [{\citenamefont {Mroue}\ \emph {et~al.}(2013)\citenamefont {Mroue},
  \citenamefont {Scheel}, \citenamefont {Szilagyi}, \citenamefont {Pfeiffer},
  \citenamefont {Boyle} \emph {et~al.}}]{Mroue:2013xna}%
  \BibitemOpen
  \bibfield  {author} {\bibinfo {author} {\bibfnamefont {A.~H.}\ \bibnamefont
  {Mroue}}, \bibinfo {author} {\bibfnamefont {M.~A.}\ \bibnamefont {Scheel}},
  \bibinfo {author} {\bibfnamefont {B.}~\bibnamefont {Szilagyi}}, \bibinfo
  {author} {\bibfnamefont {H.~P.}\ \bibnamefont {Pfeiffer}}, \bibinfo {author}
  {\bibfnamefont {M.}~\bibnamefont {Boyle}},  \emph {et~al.},\ }\href {\doibase
  10.1103/PhysRevLett.111.241104} {\bibfield  {journal} {\bibinfo  {journal}
  {Phys. Rev. Lett.}\ }\textbf {\bibinfo {volume} {111}},\ \bibinfo {pages}
  {241104} (\bibinfo {year} {2013})},\ \Eprint {http://arxiv.org/abs/1304.6077}
  {arXiv:1304.6077 [gr-qc]} \BibitemShut {NoStop}%
\bibitem [{\citenamefont {York}(1999)}]{York99}%
  \BibitemOpen
  \bibfield  {author} {\bibinfo {author} {\bibfnamefont {J.~W.}\ \bibnamefont
  {York}},\ }\href@noop {} {\bibfield  {journal} {\bibinfo  {journal} {Phys.
  Rev. Lett.}\ }\textbf {\bibinfo {volume} {82}},\ \bibinfo {pages} {1350}
  (\bibinfo {year} {1999})}\BibitemShut {NoStop}%
\bibitem [{\citenamefont {Cook}(2000)}]{Cook:2000vr}%
  \BibitemOpen
  \bibfield  {author} {\bibinfo {author} {\bibfnamefont {G.~B.}\ \bibnamefont
  {Cook}},\ }\href@noop {} {\bibfield  {journal} {\bibinfo  {journal} {Living
  Rev. Rel.}\ }\textbf {\bibinfo {volume} {3}},\ \bibinfo {pages} {5} (\bibinfo
  {year} {2000})},\ \Eprint {http://arxiv.org/abs/gr-qc/0007085}
  {arXiv:gr-qc/0007085 [gr-qc]} \BibitemShut {NoStop}%
\bibitem [{\citenamefont {Pfeiffer}\ and\ \citenamefont
  {York}(2003)}]{Pfeiffer:2002iy}%
  \BibitemOpen
  \bibfield  {author} {\bibinfo {author} {\bibfnamefont {H.~P.}\ \bibnamefont
  {Pfeiffer}}\ and\ \bibinfo {author} {\bibfnamefont {J.}~\bibnamefont {York},
  \bibfnamefont {James~W.}},\ }\href@noop {} {\bibfield  {journal} {\bibinfo
  {journal} {Phys. Rev. D}\ }\textbf {\bibinfo {volume} {67}},\ \bibinfo
  {pages} {044022} (\bibinfo {year} {2003})},\ \Eprint
  {http://arxiv.org/abs/gr-qc/0207095} {gr-qc/0207095} \BibitemShut {NoStop}%
\bibitem [{\citenamefont {{Alcubierre}}(2008)}]{AlcubierreBook2008}%
  \BibitemOpen
  \bibfield  {author} {\bibinfo {author} {\bibfnamefont {M.}~\bibnamefont
  {{Alcubierre}}},\ }\href@noop {} {\emph {\bibinfo {title} {Introduction to
  3+1 Numerical Relativity, by Miguel Alcubierre.~ISBN 978-0-19-920567-7
  (HB).~Published by Oxford University Press, Oxford, UK, 2008.}}}\ (\bibinfo
  {publisher} {Oxford University Press},\ \bibinfo {year} {2008})\BibitemShut
  {NoStop}%
\bibitem [{\citenamefont {Ansorg}\ \emph {et~al.}(2004)\citenamefont {Ansorg},
  \citenamefont {Br\"ugmann},\ and\ \citenamefont {Tichy}}]{Ansorg:2004ds}%
  \BibitemOpen
  \bibfield  {author} {\bibinfo {author} {\bibfnamefont {M.}~\bibnamefont
  {Ansorg}}, \bibinfo {author} {\bibfnamefont {B.}~\bibnamefont {Br\"ugmann}},
  \ and\ \bibinfo {author} {\bibfnamefont {W.}~\bibnamefont {Tichy}},\
  }\href@noop {} {\bibfield  {journal} {\bibinfo  {journal} {Phys. Rev.}\
  }\textbf {\bibinfo {volume} {D70}},\ \bibinfo {pages} {064011} (\bibinfo
  {year} {2004})},\ \Eprint {http://arxiv.org/abs/gr-qc/0404056}
  {gr-qc/0404056} \BibitemShut {NoStop}%
\bibitem [{\citenamefont {Healy}\ \emph
  {et~al.}(2017{\natexlab{b}})\citenamefont {Healy}, \citenamefont {Lousto},
  \citenamefont {Nakano},\ and\ \citenamefont {Zlochower}}]{Healy:2017zqj}%
  \BibitemOpen
  \bibfield  {author} {\bibinfo {author} {\bibfnamefont {J.}~\bibnamefont
  {Healy}}, \bibinfo {author} {\bibfnamefont {C.~O.}\ \bibnamefont {Lousto}},
  \bibinfo {author} {\bibfnamefont {H.}~\bibnamefont {Nakano}}, \ and\ \bibinfo
  {author} {\bibfnamefont {Y.}~\bibnamefont {Zlochower}},\ }\href {\doibase
  10.1088/1361-6382/aa7929} {\bibfield  {journal} {\bibinfo  {journal} {Class.
  Quant. Grav.}\ }\textbf {\bibinfo {volume} {34}},\ \bibinfo {pages} {145011}
  (\bibinfo {year} {2017}{\natexlab{b}})},\ \Eprint
  {http://arxiv.org/abs/1702.00872} {arXiv:1702.00872 [gr-qc]} \BibitemShut
  {NoStop}%
\bibitem [{\citenamefont {Zlochower}\ \emph {et~al.}(2005)\citenamefont
  {Zlochower}, \citenamefont {Baker}, \citenamefont {Campanelli},\ and\
  \citenamefont {Lousto}}]{Zlochower:2005bj}%
  \BibitemOpen
  \bibfield  {author} {\bibinfo {author} {\bibfnamefont {Y.}~\bibnamefont
  {Zlochower}}, \bibinfo {author} {\bibfnamefont {J.~G.}\ \bibnamefont
  {Baker}}, \bibinfo {author} {\bibfnamefont {M.}~\bibnamefont {Campanelli}}, \
  and\ \bibinfo {author} {\bibfnamefont {C.~O.}\ \bibnamefont {Lousto}},\
  }\href {\doibase 10.1103/PhysRevD.72.024021} {\bibfield  {journal} {\bibinfo
  {journal} {Phys. Rev.}\ }\textbf {\bibinfo {volume} {D72}},\ \bibinfo {pages}
  {024021} (\bibinfo {year} {2005})},\ \Eprint
  {http://arxiv.org/abs/gr-qc/0505055} {arXiv:gr-qc/0505055} \BibitemShut
  {NoStop}%
\bibitem [{\citenamefont {Alic}\ \emph {et~al.}(2012)\citenamefont {Alic},
  \citenamefont {Bona-Casas}, \citenamefont {Bona}, \citenamefont {Rezzolla},\
  and\ \citenamefont {Palenzuela}}]{Alic:2011gg}%
  \BibitemOpen
  \bibfield  {author} {\bibinfo {author} {\bibfnamefont {D.}~\bibnamefont
  {Alic}}, \bibinfo {author} {\bibfnamefont {C.}~\bibnamefont {Bona-Casas}},
  \bibinfo {author} {\bibfnamefont {C.}~\bibnamefont {Bona}}, \bibinfo {author}
  {\bibfnamefont {L.}~\bibnamefont {Rezzolla}}, \ and\ \bibinfo {author}
  {\bibfnamefont {C.}~\bibnamefont {Palenzuela}},\ }\href {\doibase
  10.1103/PhysRevD.85.064040} {\bibfield  {journal} {\bibinfo  {journal} {Phys.
  Rev.}\ }\textbf {\bibinfo {volume} {D85}},\ \bibinfo {pages} {064040}
  (\bibinfo {year} {2012})},\ \Eprint {http://arxiv.org/abs/1106.2254}
  {arXiv:1106.2254 [gr-qc]} \BibitemShut {NoStop}%
\bibitem [{\citenamefont {Nakamura}\ \emph {et~al.}(1987)\citenamefont
  {Nakamura}, \citenamefont {Oohara},\ and\ \citenamefont
  {Kojima}}]{Nakamura87}%
  \BibitemOpen
  \bibfield  {author} {\bibinfo {author} {\bibfnamefont {T.}~\bibnamefont
  {Nakamura}}, \bibinfo {author} {\bibfnamefont {K.}~\bibnamefont {Oohara}}, \
  and\ \bibinfo {author} {\bibfnamefont {Y.}~\bibnamefont {Kojima}},\
  }\href@noop {} {\bibfield  {journal} {\bibinfo  {journal} {Prog. Theor. Phys.
  Suppl.}\ }\textbf {\bibinfo {volume} {90}},\ \bibinfo {pages} {1} (\bibinfo
  {year} {1987})}\BibitemShut {NoStop}%
\bibitem [{\citenamefont {Shibata}\ and\ \citenamefont
  {Nakamura}(1995)}]{Shibata95}%
  \BibitemOpen
  \bibfield  {author} {\bibinfo {author} {\bibfnamefont {M.}~\bibnamefont
  {Shibata}}\ and\ \bibinfo {author} {\bibfnamefont {T.}~\bibnamefont
  {Nakamura}},\ }\href@noop {} {\bibfield  {journal} {\bibinfo  {journal}
  {Phys. Rev.}\ }\textbf {\bibinfo {volume} {D52}},\ \bibinfo {pages} {5428}
  (\bibinfo {year} {1995})}\BibitemShut {NoStop}%
\bibitem [{\citenamefont {Baumgarte}\ and\ \citenamefont
  {Shapiro}(1998)}]{Baumgarte99}%
  \BibitemOpen
  \bibfield  {author} {\bibinfo {author} {\bibfnamefont {T.~W.}\ \bibnamefont
  {Baumgarte}}\ and\ \bibinfo {author} {\bibfnamefont {S.~L.}\ \bibnamefont
  {Shapiro}},\ }\href@noop {} {\bibfield  {journal} {\bibinfo  {journal} {Phys.
  Rev.}\ }\textbf {\bibinfo {volume} {D59}},\ \bibinfo {pages} {024007}
  (\bibinfo {year} {1998})},\ \Eprint {http://arxiv.org/abs/gr-qc/9810065}
  {gr-qc/9810065} \BibitemShut {NoStop}%
\bibitem [{\citenamefont {Lousto}\ and\ \citenamefont
  {Zlochower}(2008)}]{Lousto:2007rj}%
  \BibitemOpen
  \bibfield  {author} {\bibinfo {author} {\bibfnamefont {C.~O.}\ \bibnamefont
  {Lousto}}\ and\ \bibinfo {author} {\bibfnamefont {Y.}~\bibnamefont
  {Zlochower}},\ }\href {\doibase 10.1103/PhysRevD.77.024034} {\bibfield
  {journal} {\bibinfo  {journal} {Phys. Rev.}\ }\textbf {\bibinfo {volume}
  {D77}},\ \bibinfo {pages} {024034} (\bibinfo {year} {2008})},\ \Eprint
  {http://arxiv.org/abs/0711.1165} {arXiv:0711.1165 [gr-qc]} \BibitemShut
  {NoStop}%
\bibitem [{cac()}]{cactus_web}%
  \BibitemOpen
  \href@noop {} {}\bibinfo {note} {Cactus Computational Toolkit home page: {\tt
  http://cactuscode.org}}\BibitemShut {NoStop}%
\bibitem [{ein()}]{einsteintoolkit}%
  \BibitemOpen
  \href@noop {} {}\bibinfo {note} {Einstein Toolkit home page: {\tt
  http://einsteintoolkit.org}}\BibitemShut {NoStop}%
\bibitem [{\citenamefont {Schnetter}\ \emph {et~al.}(2004)\citenamefont
  {Schnetter}, \citenamefont {Hawley},\ and\ \citenamefont
  {Hawke}}]{Schnetter-etal-03b}%
  \BibitemOpen
  \bibfield  {author} {\bibinfo {author} {\bibfnamefont {E.}~\bibnamefont
  {Schnetter}}, \bibinfo {author} {\bibfnamefont {S.~H.}\ \bibnamefont
  {Hawley}}, \ and\ \bibinfo {author} {\bibfnamefont {I.}~\bibnamefont
  {Hawke}},\ }\href@noop {} {\bibfield  {journal} {\bibinfo  {journal} {Class.
  Quant. Grav.}\ }\textbf {\bibinfo {volume} {21}},\ \bibinfo {pages} {1465}
  (\bibinfo {year} {2004})},\ \Eprint {http://arxiv.org/abs/gr-qc/0310042}
  {gr-qc/0310042} \BibitemShut {NoStop}%
\bibitem [{\citenamefont {Thornburg}(2004)}]{Thornburg2003:AH-finding}%
  \BibitemOpen
  \bibfield  {author} {\bibinfo {author} {\bibfnamefont {J.}~\bibnamefont
  {Thornburg}},\ }\href {\doibase 10.1088/0264-9381/21/2/026} {\bibfield
  {journal} {\bibinfo  {journal} {Class. Quant. Grav.}\ }\textbf {\bibinfo
  {volume} {21}},\ \bibinfo {pages} {743} (\bibinfo {year} {2004})},\ \Eprint
  {http://arxiv.org/abs/gr-qc/0306056} {gr-qc/0306056} \BibitemShut {NoStop}%
\bibitem [{\citenamefont {Dreyer}\ \emph {et~al.}(2003)\citenamefont {Dreyer},
  \citenamefont {Krishnan}, \citenamefont {Shoemaker},\ and\ \citenamefont
  {Schnetter}}]{Dreyer02a}%
  \BibitemOpen
  \bibfield  {author} {\bibinfo {author} {\bibfnamefont {O.}~\bibnamefont
  {Dreyer}}, \bibinfo {author} {\bibfnamefont {B.}~\bibnamefont {Krishnan}},
  \bibinfo {author} {\bibfnamefont {D.}~\bibnamefont {Shoemaker}}, \ and\
  \bibinfo {author} {\bibfnamefont {E.}~\bibnamefont {Schnetter}},\ }\href@noop
  {} {\bibfield  {journal} {\bibinfo  {journal} {Phys. Rev.}\ }\textbf
  {\bibinfo {volume} {D67}},\ \bibinfo {pages} {024018} (\bibinfo {year}
  {2003})},\ \Eprint {http://arxiv.org/abs/gr-qc/0206008} {gr-qc/0206008}
  \BibitemShut {NoStop}%
\bibitem [{\citenamefont {Campanelli}\ \emph
  {et~al.}(2006{\natexlab{c}})\citenamefont {Campanelli}, \citenamefont
  {Kelly},\ and\ \citenamefont {Lousto}}]{Campanelli:2005ia}%
  \BibitemOpen
  \bibfield  {author} {\bibinfo {author} {\bibfnamefont {M.}~\bibnamefont
  {Campanelli}}, \bibinfo {author} {\bibfnamefont {B.~J.}\ \bibnamefont
  {Kelly}}, \ and\ \bibinfo {author} {\bibfnamefont {C.~O.}\ \bibnamefont
  {Lousto}},\ }\href {\doibase 10.1103/PhysRevD.73.064005} {\bibfield
  {journal} {\bibinfo  {journal} {Phys. Rev.}\ }\textbf {\bibinfo {volume}
  {D73}},\ \bibinfo {pages} {064005} (\bibinfo {year} {2006}{\natexlab{c}})},\
  \Eprint {http://arxiv.org/abs/gr-qc/0510122} {arXiv:gr-qc/0510122}
  \BibitemShut {NoStop}%
\bibitem [{\citenamefont {Baker}\ \emph {et~al.}(2002)\citenamefont {Baker},
  \citenamefont {Campanelli},\ and\ \citenamefont {Lousto}}]{Baker:2001sf}%
  \BibitemOpen
  \bibfield  {author} {\bibinfo {author} {\bibfnamefont {J.~G.}\ \bibnamefont
  {Baker}}, \bibinfo {author} {\bibfnamefont {M.}~\bibnamefont {Campanelli}}, \
  and\ \bibinfo {author} {\bibfnamefont {C.~O.}\ \bibnamefont {Lousto}},\
  }\href {\doibase 10.1103/PhysRevD.65.044001} {\bibfield  {journal} {\bibinfo
  {journal} {Phys. Rev.}\ }\textbf {\bibinfo {volume} {D65}},\ \bibinfo {pages}
  {044001} (\bibinfo {year} {2002})},\ \Eprint
  {http://arxiv.org/abs/gr-qc/0104063} {arXiv:gr-qc/0104063 [gr-qc]}
  \BibitemShut {NoStop}%
\bibitem [{\citenamefont {Nakano}\ \emph {et~al.}(2015)\citenamefont {Nakano},
  \citenamefont {Healy}, \citenamefont {Lousto},\ and\ \citenamefont
  {Zlochower}}]{Nakano:2015pta}%
  \BibitemOpen
  \bibfield  {author} {\bibinfo {author} {\bibfnamefont {H.}~\bibnamefont
  {Nakano}}, \bibinfo {author} {\bibfnamefont {J.}~\bibnamefont {Healy}},
  \bibinfo {author} {\bibfnamefont {C.~O.}\ \bibnamefont {Lousto}}, \ and\
  \bibinfo {author} {\bibfnamefont {Y.}~\bibnamefont {Zlochower}},\ }\href
  {\doibase 10.1103/PhysRevD.91.104022} {\bibfield  {journal} {\bibinfo
  {journal} {Phys. Rev.}\ }\textbf {\bibinfo {volume} {D91}},\ \bibinfo {pages}
  {104022} (\bibinfo {year} {2015})},\ \Eprint
  {http://arxiv.org/abs/1503.00718} {arXiv:1503.00718 [gr-qc]} \BibitemShut
  {NoStop}%
\bibitem [{\citenamefont {Alcubierre}\ \emph {et~al.}(2003)\citenamefont
  {Alcubierre}, \citenamefont {Br\"ugmann}, \citenamefont {Diener},
  \citenamefont {Koppitz}, \citenamefont {Pollney}, \citenamefont {Seidel},\
  and\ \citenamefont {Takahashi}}]{Alcubierre02a}%
  \BibitemOpen
  \bibfield  {author} {\bibinfo {author} {\bibfnamefont {M.}~\bibnamefont
  {Alcubierre}}, \bibinfo {author} {\bibfnamefont {B.}~\bibnamefont
  {Br\"ugmann}}, \bibinfo {author} {\bibfnamefont {P.}~\bibnamefont {Diener}},
  \bibinfo {author} {\bibfnamefont {M.}~\bibnamefont {Koppitz}}, \bibinfo
  {author} {\bibfnamefont {D.}~\bibnamefont {Pollney}}, \bibinfo {author}
  {\bibfnamefont {E.}~\bibnamefont {Seidel}}, \ and\ \bibinfo {author}
  {\bibfnamefont {R.}~\bibnamefont {Takahashi}},\ }\href@noop {} {\bibfield
  {journal} {\bibinfo  {journal} {Phys. Rev.}\ }\textbf {\bibinfo {volume}
  {D67}},\ \bibinfo {pages} {084023} (\bibinfo {year} {2003})},\ \Eprint
  {http://arxiv.org/abs/gr-qc/0206072} {gr-qc/0206072} \BibitemShut {NoStop}%
\bibitem [{\citenamefont {van Meter}\ \emph {et~al.}(2006)\citenamefont {van
  Meter}, \citenamefont {Baker}, \citenamefont {Koppitz},\ and\ \citenamefont
  {Choi}}]{vanMeter:2006vi}%
  \BibitemOpen
  \bibfield  {author} {\bibinfo {author} {\bibfnamefont {J.~R.}\ \bibnamefont
  {van Meter}}, \bibinfo {author} {\bibfnamefont {J.~G.}\ \bibnamefont
  {Baker}}, \bibinfo {author} {\bibfnamefont {M.}~\bibnamefont {Koppitz}}, \
  and\ \bibinfo {author} {\bibfnamefont {D.-I.}\ \bibnamefont {Choi}},\
  }\href@noop {} {\bibfield  {journal} {\bibinfo  {journal} {Phys. Rev.}\
  }\textbf {\bibinfo {volume} {D73}},\ \bibinfo {pages} {124011} (\bibinfo
  {year} {2006})},\ \Eprint {http://arxiv.org/abs/gr-qc/0605030}
  {gr-qc/0605030} \BibitemShut {NoStop}%
\bibitem [{\citenamefont {Schnetter}(2010)}]{Schnetter:2010cz}%
  \BibitemOpen
  \bibfield  {author} {\bibinfo {author} {\bibfnamefont {E.}~\bibnamefont
  {Schnetter}},\ }\href {\doibase 10.1088/0264-9381/27/16/167001} {\bibfield
  {journal} {\bibinfo  {journal} {Class. Quant. Grav.}\ }\textbf {\bibinfo
  {volume} {27}},\ \bibinfo {pages} {167001} (\bibinfo {year} {2010})},\
  \Eprint {http://arxiv.org/abs/1003.0859} {arXiv:1003.0859 [gr-qc]}
  \BibitemShut {NoStop}%
\bibitem [{\citenamefont {Pfeiffer}\ \emph {et~al.}(2007)\citenamefont
  {Pfeiffer}, \citenamefont {Brown}, \citenamefont {Kidder}, \citenamefont
  {Lindblom}, \citenamefont {Lovelace},\ and\ \citenamefont
  {Scheel}}]{Pfeiffer:2007yz}%
  \BibitemOpen
  \bibfield  {author} {\bibinfo {author} {\bibfnamefont {H.~P.}\ \bibnamefont
  {Pfeiffer}}, \bibinfo {author} {\bibfnamefont {D.~A.}\ \bibnamefont {Brown}},
  \bibinfo {author} {\bibfnamefont {L.~E.}\ \bibnamefont {Kidder}}, \bibinfo
  {author} {\bibfnamefont {L.}~\bibnamefont {Lindblom}}, \bibinfo {author}
  {\bibfnamefont {G.}~\bibnamefont {Lovelace}}, \ and\ \bibinfo {author}
  {\bibfnamefont {M.~A.}\ \bibnamefont {Scheel}},\ }\href {\doibase
  10.1088/0264-9381/24/12/S06} {\bibfield  {journal} {\bibinfo  {journal}
  {Class. Quant. Grav.}\ }\textbf {\bibinfo {volume} {24}},\ \bibinfo {pages}
  {S59} (\bibinfo {year} {2007})},\ \Eprint
  {http://arxiv.org/abs/gr-qc/0702106} {arXiv:gr-qc/0702106 [gr-qc]}
  \BibitemShut {NoStop}%
\bibitem [{\citenamefont {Buonanno}\ \emph {et~al.}(2011)\citenamefont
  {Buonanno}, \citenamefont {Kidder}, \citenamefont {Mroue}, \citenamefont
  {Pfeiffer},\ and\ \citenamefont {Taracchini}}]{Buonanno:2010yk}%
  \BibitemOpen
  \bibfield  {author} {\bibinfo {author} {\bibfnamefont {A.}~\bibnamefont
  {Buonanno}}, \bibinfo {author} {\bibfnamefont {L.~E.}\ \bibnamefont
  {Kidder}}, \bibinfo {author} {\bibfnamefont {A.~H.}\ \bibnamefont {Mroue}},
  \bibinfo {author} {\bibfnamefont {H.~P.}\ \bibnamefont {Pfeiffer}}, \ and\
  \bibinfo {author} {\bibfnamefont {A.}~\bibnamefont {Taracchini}},\ }\href
  {\doibase 10.1103/PhysRevD.83.104034} {\bibfield  {journal} {\bibinfo
  {journal} {Phys. Rev.}\ }\textbf {\bibinfo {volume} {D83}},\ \bibinfo {pages}
  {104034} (\bibinfo {year} {2011})},\ \Eprint {http://arxiv.org/abs/1012.1549}
  {arXiv:1012.1549 [gr-qc]} \BibitemShut {NoStop}%
\bibitem [{\citenamefont {Purrer}\ \emph {et~al.}(2012)\citenamefont {Purrer},
  \citenamefont {Husa},\ and\ \citenamefont {Hannam}}]{Purrer:2012wy}%
  \BibitemOpen
  \bibfield  {author} {\bibinfo {author} {\bibfnamefont {M.}~\bibnamefont
  {Purrer}}, \bibinfo {author} {\bibfnamefont {S.}~\bibnamefont {Husa}}, \ and\
  \bibinfo {author} {\bibfnamefont {M.}~\bibnamefont {Hannam}},\ }\href
  {\doibase 10.1103/PhysRevD.85.124051} {\bibfield  {journal} {\bibinfo
  {journal} {Phys. Rev.}\ }\textbf {\bibinfo {volume} {D85}},\ \bibinfo {pages}
  {124051} (\bibinfo {year} {2012})},\ \Eprint {http://arxiv.org/abs/1203.4258}
  {arXiv:1203.4258 [gr-qc]} \BibitemShut {NoStop}%
\bibitem [{\citenamefont {Buchman}\ \emph {et~al.}(2012)\citenamefont
  {Buchman}, \citenamefont {Pfeiffer}, \citenamefont {Scheel},\ and\
  \citenamefont {Szilagyi}}]{Buchman:2012dw}%
  \BibitemOpen
  \bibfield  {author} {\bibinfo {author} {\bibfnamefont {L.~T.}\ \bibnamefont
  {Buchman}}, \bibinfo {author} {\bibfnamefont {H.~P.}\ \bibnamefont
  {Pfeiffer}}, \bibinfo {author} {\bibfnamefont {M.~A.}\ \bibnamefont
  {Scheel}}, \ and\ \bibinfo {author} {\bibfnamefont {B.}~\bibnamefont
  {Szilagyi}},\ }\href {\doibase 10.1103/PhysRevD.86.084033} {\bibfield
  {journal} {\bibinfo  {journal} {Phys. Rev.}\ }\textbf {\bibinfo {volume}
  {D86}},\ \bibinfo {pages} {084033} (\bibinfo {year} {2012})},\ \Eprint
  {http://arxiv.org/abs/1206.3015} {arXiv:1206.3015 [gr-qc]} \BibitemShut
  {NoStop}%
\bibitem [{\citenamefont {Campanelli}\ and\ \citenamefont
  {Lousto}(1999)}]{Campanelli:1998jv}%
  \BibitemOpen
  \bibfield  {author} {\bibinfo {author} {\bibfnamefont {M.}~\bibnamefont
  {Campanelli}}\ and\ \bibinfo {author} {\bibfnamefont {C.~O.}\ \bibnamefont
  {Lousto}},\ }\href {\doibase 10.1103/PhysRevD.59.124022} {\bibfield
  {journal} {\bibinfo  {journal} {Phys. Rev.}\ }\textbf {\bibinfo {volume}
  {D59}},\ \bibinfo {pages} {124022} (\bibinfo {year} {1999})},\ \Eprint
  {http://arxiv.org/abs/gr-qc/9811019} {arXiv:gr-qc/9811019 [gr-qc]}
  \BibitemShut {NoStop}%
\bibitem [{\citenamefont {Lousto}\ and\ \citenamefont
  {Zlochower}(2007)}]{Lousto:2007mh}%
  \BibitemOpen
  \bibfield  {author} {\bibinfo {author} {\bibfnamefont {C.~O.}\ \bibnamefont
  {Lousto}}\ and\ \bibinfo {author} {\bibfnamefont {Y.}~\bibnamefont
  {Zlochower}},\ }\href@noop {} {\bibfield  {journal} {\bibinfo  {journal}
  {Phys. Rev.}\ }\textbf {\bibinfo {volume} {D76}},\ \bibinfo {pages}
  {041502(R)} (\bibinfo {year} {2007})},\ \Eprint
  {http://arxiv.org/abs/gr-qc/0703061} {gr-qc/0703061} \BibitemShut {NoStop}%
\bibitem [{\citenamefont {Lovelace}\ \emph {et~al.}(2016)\citenamefont
  {Lovelace} \emph {et~al.}}]{Lovelace:2016uwp}%
  \BibitemOpen
  \bibfield  {author} {\bibinfo {author} {\bibfnamefont {G.}~\bibnamefont
  {Lovelace}} \emph {et~al.},\ }\href {\doibase 10.1088/0264-9381/33/24/244002}
  {\bibfield  {journal} {\bibinfo  {journal} {Class. Quant. Grav.}\ }\textbf
  {\bibinfo {volume} {33}},\ \bibinfo {pages} {244002} (\bibinfo {year}
  {2016})},\ \Eprint {http://arxiv.org/abs/1607.05377} {arXiv:1607.05377
  [gr-qc]} \BibitemShut {NoStop}%
\end{thebibliography}%

\end{document}